%% file: bare_jrnl_new_sample4.tex
\documentclass[lettersize,journal]{IEEEtran}
\usepackage{tikz}
\usepackage{amsmath}
\usepackage{filecontents}

\DeclareMathOperator*{\argmax}{arg\,max} 
\DeclareMathOperator*{\argmin}{arg\,min}
\usepackage{rotating}
\usepackage{tcolorbox}
\usepackage{enumitem}
\usepackage{lipsum}
\usepackage[linesnumbered,ruled,vlined]{algorithm2e}
\usepackage{tcolorbox}
\usepackage{tabularx}
\usepackage{balance}
\usepackage{graphicx}
\usepackage{url}
\usepackage{booktabs}
\usepackage{amssymb}
\usepackage{makecell}
\usepackage[normalem]{ulem}
\usepackage{float}
\usepackage{subcaption}
\usepackage{bbm}
\usepackage{dsfont}
\usepackage{marvosym}
\usepackage{amsfonts}
\usepackage{algpseudocode}
\usepackage{bm}
\usepackage[utf8]{inputenc}
\usepackage{microtype}
\usepackage{adjustbox}
\usepackage{siunitx}
\usepackage{multirow}
\usepackage{amsthm,amsmath}
\usepackage{caption}
\usepackage{xcolor}
\usepackage{hyperref}
\usepackage{soul}
\usepackage{colortbl}  
\usepackage{xurl}
\usepackage{xcolor, eucal}
\usepackage{array}   
\usepackage{filecontents}
\usepackage[table]{xcolor}

\newcolumntype{C}{>{\columncolor{lightgray!30}}S}
\definecolor{lightgray}{gray}{0.9}

\begin{document}

\title{TooBadRL: Trigger Optimization to Boost Effectiveness of Backdoor Attacks on Deep Reinforcement Learning}

\author{
       Mingxuan Zhang, 
       Oubo Ma,
       Kang Wei,~\IEEEmembership{Member,~IEEE,}
       Songze Li,~\IEEEmembership{Member,~IEEE,} 
       and Shouling Ji,~\IEEEmembership{Senior Member,~IEEE}
       
\IEEEcompsocitemizethanks{

\IEEEcompsocthanksitem M. Zhang, K. Wei and S. Li are with Southeast University, China (E-mail: mingxuan.zhang, kang.wei, and songzeli@seu.edu.cn).

\IEEEcompsocthanksitem O. Ma, and L. Ji are with Zhejiang University, China (E-mail: mob, sji@zju.edu.cn).

%

}
}


\maketitle

\begin{abstract}
Deep reinforcement learning (DRL) has achieved remarkable success in a wide range of sequential decision-making applications, including robotics, healthcare, smart grids, and finance.
Recent studies reveal that adversaries can implant backdoors into DRL agents during the training phase. 
These backdoors can later be activated by specific triggers during deployment, compelling the agent to execute targeted actions and potentially leading to severe consequences, such as drone crashes or vehicle collisions.
However, existing backdoor attacks utilize simplistic and heuristic trigger configurations, overlooking the critical impact of trigger design on attack effectiveness.
To address this gap, we introduce \texttt{TooBadRL}, the first framework to systematically optimize DRL backdoor triggers across three critical aspects: injection timing, trigger dimension, and manipulation magnitude.
Specifically, we first introduce a performance-aware adaptive freezing mechanism to determine the injection timing during training.
Then, we formulate trigger selection as an influence attribution problem and apply Shapley value analysis to identify the most influential trigger dimension for injection.
Furthermore, we propose an adversarial input synthesis method to optimize the manipulation magnitude under environmental constraints.
Extensive evaluations on three DRL algorithms and nine benchmark tasks demonstrate that \texttt{TooBadRL} outperforms five baseline methods in terms of attack success rate while only slightly affecting normal task performance.
We further evaluate potential defense strategies from detection and mitigation perspectives.
We open-source our code to facilitate reproducibility and further research.
\end{abstract}

\begin{IEEEkeywords}
Backdoor attack, deep reinforcement learning, trigger optimization.
\end{IEEEkeywords}

\input{1-Introduction}
\input{2-Related_Work}
\input{3-DRL_Backdoor_Attack}
\input{4-Trigger_Optimization_of_TooBadRL}

\input{5-Backdoor_Implantation_of_TooBadRL}

\input{6-Evaluation}
\input{7-Conclusion}

\bibliographystyle{plain}
\bibliography{reference}

\clearpage
\input{Appendix}





\end{document}

%% file: 1-Introduction.tex
\section{Introduction}
\label{sec:intro}

\IEEEPARstart{D}EEP Reinforcement Learning (DRL) has emerged as a powerful paradigm for solving complex sequential decision-making problems.
Its success spans a wide array of domains, including large language models~\cite{NEURIPS2022_b1efde53}, finance~\cite{Deng2017Deep}, autonomous driving~\cite{Sun2023Hierarchical}, and healthcare~\cite{jumper2021highly}. 
However, as DRL systems are increasingly deployed in safety-critical applications, their reliability and security have become paramount concerns~\cite{Wu2021Adversarial, SUB2024Oubo, RU2023Dayong, du2025revealing}. 
Among various security threats, backdoor attacks~\cite{Xuan2025Temporal, Yang2025TrojanTO} pose particularly severe risks to the trustworthiness of DRL systems.

In a backdoor attack, an adversary manipulates the training process to establish a hidden association between a specific trigger pattern and a predetermined malicious action within the DRL agent's policy~\cite{Yao2019Latent,kiourti2020trojdrl,cui2024badrl}.
The backdoored agent behaves normally on clean inputs but executes the malicious action when the trigger is present.
For example, in robotic manipulation, triggers could cause robots to drop components or strike nearby objects during assembly tasks~\cite{Wang2024TrojanRobot}.
Such backdoor-triggered failures cause equipment damage and threaten human safety, undermining safe DRL deployment in critical applications~\cite{Xuan2025Temporal, Han2022BadDet}.

While existing studies have explored DRL backdoor attacks~\cite{ma2025unidooruniversalframeworkactionlevel,rathbun2024sleepernets,yang2019design}, most methods rely on heuristic trigger design without any methodological guidance.
For instance, TrojDRL~\cite{kiourti2020trojdrl} injects triggers from the training start, randomly selects state dimensions, and assigns boundary values as triggers.
Similarly, existing approaches commonly use fixed freezing period ratios for injection timing~\cite{ma2025unidooruniversalframeworkactionlevel}, while BAFFLE~\cite{gong2024baffle} arbitrarily selects dimensions and assigns fixed midpoint values as magnitudes.
These heuristic configurations lack principled justification, leaving a critical question unanswered: how does trigger design impact backdoor attacks?

To address the above critical question, we decompose trigger design into three critical aspects: injection timing (when to begin injecting triggers during training), trigger dimension (which state dimensions to perturb), and manipulation magnitude (what values to assign to perturbed dimensions).
We conduct a preliminary controlled experiment using TrojDRL~\cite{kiourti2020trojdrl} as the baseline to quantify how the three critical aspects impact attack performance.
We independently vary each aspect while holding others constant: injection timing ranges from $0\%$ to $100\%$ of training progress, trigger dimension covers each individual state dimension, and manipulation magnitude spans the valid range for the selected dimension.

\begin{figure}[!t]
    \centering
    \begin{subfigure}{0.6\linewidth}
        \centering
        \includegraphics[width=\linewidth]{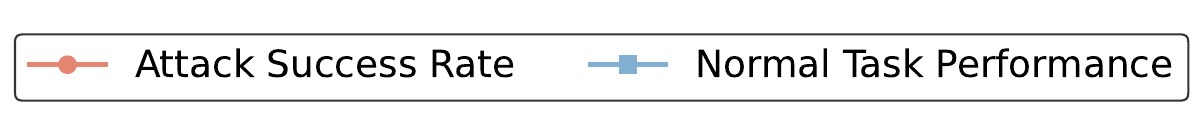}
    \end{subfigure}
    
    \begin{subfigure}{0.32\linewidth}
        \centering
        \includegraphics[width=\linewidth]{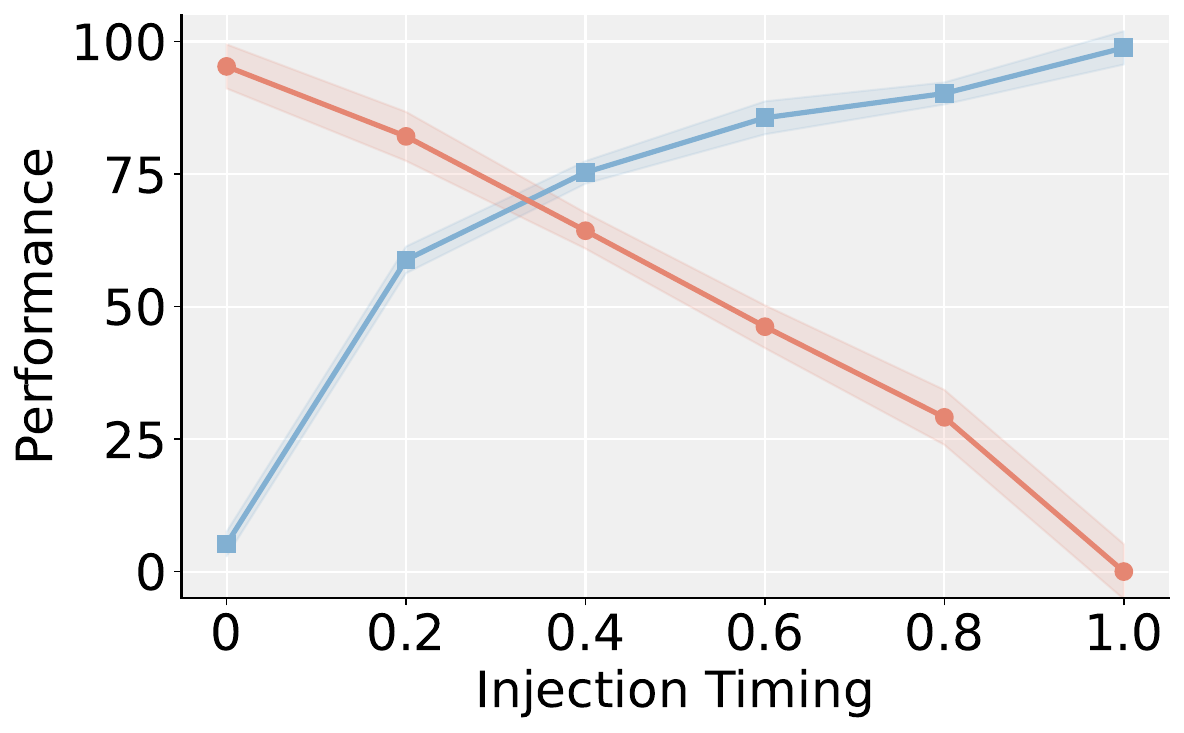}
        \caption{Timing}
        \label{fig:Figure1_Timing}
    \end{subfigure}
    \hfill 
    \begin{subfigure}{0.32\linewidth}
        \centering
        \includegraphics[width=\linewidth]{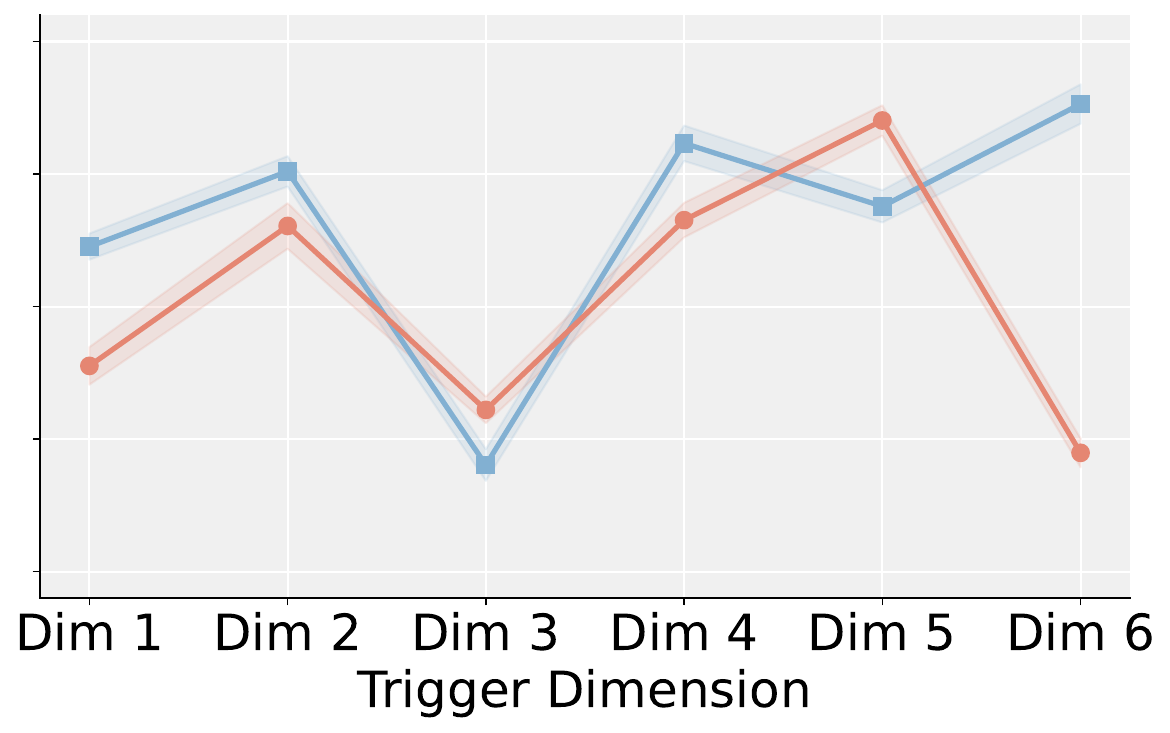}
        \caption{Dimension}
        \label{fig:Figure1_Dimension}
    \end{subfigure}
    \hfill 
    \begin{subfigure}{0.32\linewidth}
        \centering
        \includegraphics[width=\linewidth]{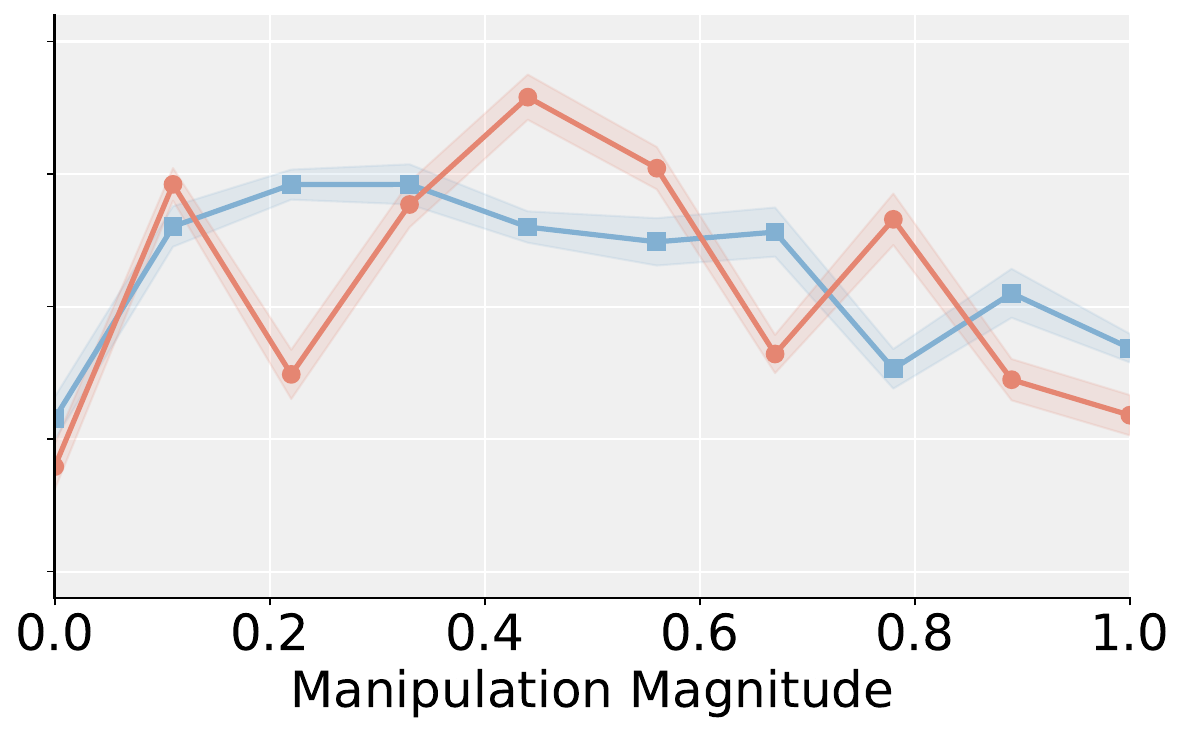}
        \caption{Magnitude}
        \label{fig:Figure1_Value}
    \end{subfigure}
    \caption{Impact of trigger configuration on attack performance. Each subfigure illustrates how varying one critical aspect affects both \textit{ASR} and \textit{NTP}.}
    \label{fig:pre}
\end{figure}

The results, as depicted in Figure~\ref{fig:pre}, reveal substantial performance volatility across the three critical aspects.
Varying injection timing causes Attack Success Rate (\textit{ASR}) and Normal Task Performance (\textit{NTP}) to exhibit maximum differences of $83\%$ and $98\%$, respectively, across environments.
Varying trigger dimension causes \textit{ASR} and \textit{NTP} to exhibit maximum differences of $64\%$ and $52\%$, respectively.
Varying manipulation magnitude causes \textit{ASR} and \textit{NTP} to exhibit maximum differences of $76\%$ and $43\%$, respectively.
These findings demonstrate that each of the three critical aspects significantly influences attack effectiveness.

Based on the above findings, we introduce \texttt{TooBadRL}, a framework that systematically optimizes backdoor triggers across the three critical aspects.
(1) \textbf{Injection Timing.}
We propose a performance-aware adaptive freezing mechanism to determine the injection timing.
This mechanism uses statistical hypothesis testing to monitor the agent's learning progress and identifies when to begin trigger injection, ensuring that injection starts only after the normal task policy has stabilized.
(2) \textbf{Trigger Dimension.}
We formulate trigger dimension selection as an influence attribution problem and employ Shapley value analysis to quantify each dimension's contribution to policy outputs.
This game-theoretic approach identifies which dimensions are most influential for establishing trigger-action associations.
(3) \textbf{Manipulation Magnitude.}
Once the trigger dimension is selected, we develop a gradient-based method to optimize manipulation magnitude.
Leveraging policy differentiability, this method iteratively adjusts trigger values to maximize the probability of inducing the target action while respecting environmental constraints.

We evaluate \texttt{TooBadRL} on three mainstream DRL algorithms across nine environments, comparing against five baseline methods.
\texttt{TooBadRL} achieves an average $24\%$ improvement in \textit{ASR} over baselines while maintaining \textit{NTP} above $90\%$.
Ablation studies confirm that each component contributes to attack effectiveness.
We further evaluate defense strategies from detection and mitigation perspectives.

In summary, our contributions are as follows:
\begin{itemize}
    \item To the best of our knowledge, \texttt{TooBadRL} is the first systematic optimization framework for DRL backdoor triggers that addresses the limitations of heuristic trigger design.
    \item We develop a Shapley value-based dimension selection method to identify which state dimensions are most influential for trigger injection. We further introduce an adaptive freezing mechanism to automatically determine when to begin trigger injection.
    \item We conduct extensive evaluations demonstrating that \texttt{TooBadRL} consistently outperforms baseline methods across diverse tasks.
    We also open-source our code to facilitate future research.
\end{itemize}

%% file: 2-Related_Work.tex
 \section{Related Work}
\label{sec:rel_work}

Backdoor attacks have attracted significant attention across various deep learning domains~\cite{Lin2020Composite,yang2021rethinking,severi2021explanation,Zhang2021Graph,saha2022backdoor,bagdasaryan2020backdoor}. 
As DRL systems are increasingly deployed in safety-critical applications, these systems inherit similar vulnerabilities.
Backdoor attacks pose significant threats to both training and deployment safety of DRL agents~\cite{yuan2024shine,wang2021stop,chen2023bird,yu2024spatiotemporal}.

DRL backdoor attacks have been explored across diverse settings.
In offline RL, BAFFLE demonstrates that poisoning a small fraction of trajectories can degrade agent performance~\cite{gong2024baffle}.
In online RL, BACKDOORL targets competitive multi-agent scenarios~\cite{wang2021backdoorlbackdoorattackcompetitive}, while MARNet focuses on cooperative multi-agent systems~\cite{9894692}.
Recent work has also investigated temporal-pattern triggers for partially observable environments~\cite{yu2022temporal} and backdoor attacks under distributed learning settings~\cite{ma2025unidooruniversalframeworkactionlevel,rathbun2024sleepernets}.
These studies demonstrate the feasibility of backdoor attacks across diverse DRL contexts.

\begin{figure*}[!t]
 \centering
 \includegraphics[width=0.90\linewidth]{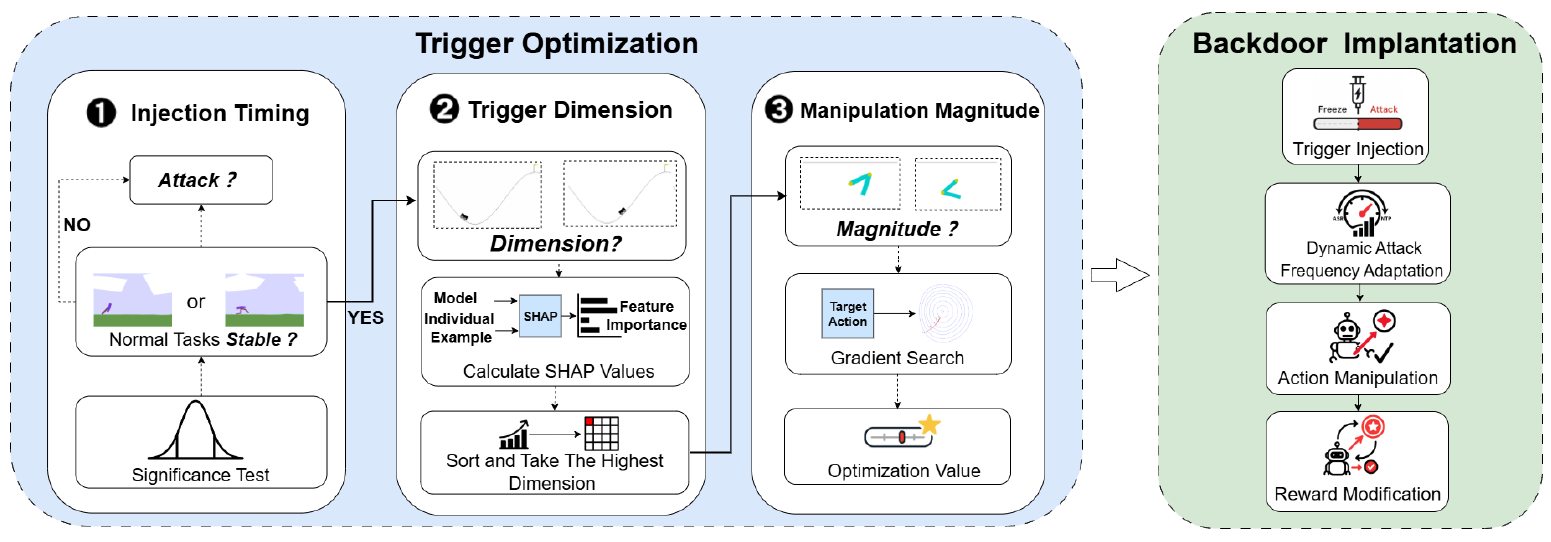}
 \caption{The framework of \texttt{TooBadRL}. \texttt{TooBadRL} operates through two stages: \textbf{(1) Trigger Optimization} systematically optimizes injection timing, trigger dimension, and manipulation magnitude; \textbf{(2) Backdoor Implantation} embeds the backdoor through trigger injection, dynamic attack frequency adaptation, action manipulation, and reward modification.}
 \label{fig:5.1-overview.png}
\end{figure*}

Despite this diversity, most existing methods rely on heuristic trigger design.
TrojDRL~\cite{kiourti2020trojdrl} injects triggers from the training start, randomly selects state dimensions, and assigns boundary values as triggers.
UNIDOOR~\cite{ma2025unidooruniversalframeworkactionlevel} uses simple threshold values despite its adaptive reward tuning mechanism.
BAFFLE~\cite{gong2024baffle} selects trigger dimensions arbitrarily and assigns fixed midpoint values.
These heuristic configurations produce inconsistent attack performance across diverse environments.

Some recent work has begun exploring optimization of individual aspects.
BadRL~\cite{cui2024badrl} leverages mutual information to select trigger dimensions, representing an initial step toward data-driven trigger design.
However, this work focuses only on the trigger dimension.

To the best of our knowledge, \texttt{TooBadRL} is the first framework to systematically optimize all three critical aspects of trigger design: injection timing, trigger dimension, and manipulation magnitude.

%% file: 3-DRL_Backdoor_Attack.tex
\section{DRL Backdoor Attack}
\label{sec:problem_formulation}

\subsection{Deep Reinforcement Learning}
\label{subsec:DRL}

DRL is typically modeled as a Markov Decision Process (MDP), defined by a tuple $\mathcal{M} = (\mathcal{S}, \mathcal{A}, R, T, \gamma)$, where $\mathcal{S}$ is the state space, $\mathcal{A}$ is the action space, $R: \mathcal{S} \times \mathcal{A} \rightarrow \mathbb{R}$ is the reward function, $T: \mathcal{S} \times \mathcal{A} \rightarrow \mathcal{S}$ is the state transition function, and $\gamma \in [0,1)$ is the discount factor.

At each timestep $t$, the agent observes a state $s_t \in \mathcal{S}$ and selects an action $a_t \in \mathcal{A}$ according to its policy $\pi_\theta(a_t|s_t)$, parameterized by neural network weights $\theta$. 
The environment provides a reward $r_t = R(s_t, a_t)$ and transitions to a new state $s_{t+1}$. 
A sequence of interactions generates a trajectory $\tau = \{(s_0, a_0, r_0), (s_1, a_1, r_1), \ldots, (s_T, a_T, r_T)\}$, where $T$ denotes the episode length. 
The agent's objective is to learn optimal policy parameters $\theta^*$ that maximize the expected cumulative discounted reward:
\begin{equation}\label{eq:drl_objective_problem}
\theta^* = \arg\max_{\theta} \mathbb{E}_{\tau \sim \pi_\theta} \left[ \sum_{t=0}^{T} \gamma^t r_t \right].
\end{equation}

The value function $V_\pi(s)$ represents the expected cumulative reward when starting from state $s$ and following policy $\pi$:
\begin{equation}\label{eq:value_function}
V_\pi(s) = \mathbb{E}_{\tau \sim \pi} \left[ \sum_{t=0}^{T} \gamma^t r_t \mid s_0 = s \right].
\end{equation}

DRL uses deep neural networks to approximate $\pi_\theta$ and value functions, making it suitable for complex state spaces~\cite{yu2018crafting, hodge2021deep, wang2019autonomous}.

\subsection{Threat Model}
\label{subsec:threat_model}

\noindent\textbf{Adversary's Objective.}
The adversary seeks to implant a hidden backdoor into a DRL agent during training.
The backdoor must satisfy two objectives: when triggers are present, the agent executes a predetermined malicious action; when triggers are absent, the agent maintains its sequential decision-making capability on normal tasks.
These two objectives create inherent tension: aggressive trigger configurations may enable reliable backdoor activation but degrade normal task performance and risk detection, while conservative configurations may preserve normal task performance but fail to reliably activate the backdoor.
The challenge lies in systematically designing triggers that balance both objectives.

\noindent\textbf{Adversary's Capabilities and Knowledge.}
We consider a supply chain attack scenario where the adversary controls the training process before deployment.
This represents either a malicious model provider who trains and distributes backdoored agents, or an adversary who obtains target agents from third-party model-sharing platforms and retrains them with backdoors before redistribution.
During training, the adversary can manipulate training signals, such as rewards and collected transitions, to embed the backdoor.
After deployment, the adversary loses access to the agent's internal parameters but can activate the backdoor by injecting predefined triggers into the environment or observations.
This threat model aligns with existing DRL backdoor attacks~\cite{kiourti2020trojdrl,cui2024badrl,ma2025unidooruniversalframeworkactionlevel,gong2024baffle} and reflects realistic security concerns in scenarios where users deploy agents from untrusted sources.

\subsection{Problem Formulation}
\label{subsec:problem_statement}
The core of backdoor attacks lies in establishing a strong association between the trigger pattern and the target action during training. 
Given a $D$-dimensional state $s = [s_1, s_2, \ldots, s_D]$, the adversary designs a trigger by selecting specific dimensions $\mathcal{I}_{\text{trigger}} \subseteq \{1, 2, \ldots, D\}$ and assigning them manipulation magnitudes. 
The triggered state $s'$ is defined as:
\begin{equation}\label{eq:trigger_definition_problem}
s'_j =
\begin{cases}
v_{\text{trigger}, j}, & \text{if } j \in \mathcal{I}_{\text{trigger}} \\
s_j, & \text{otherwise}
\end{cases}
\end{equation}
where $v_{\text{trigger}, j}$ denotes the manipulation magnitude assigned to dimension $j$, and $s'_j$ represents the $j$-th dimension of the triggered state. 
We use $v_{\text{trigger}, j}^*$ to denote the optimized manipulation magnitude obtained through gradient-based optimization (Section~\ref{subsec:trigger_value}).

Let $\pi_{\theta^*}$ denote the optimal policy learned by a clean agent that maximizes expected rewards in MDP $\mathcal{M}$ according to Eq.~\ref{eq:drl_objective_problem}. 
The adversary aims to train a backdoored policy $\pi_{\theta_b}$, parameterized by weights $\theta_b$, that satisfies two objectives:
\begin{itemize}[leftmargin=*]
    \item \textbf{Attack Effectiveness:} Maximize the probability that $\pi_{\theta_b}$ selects the target action $a_{\text{target}}$ when given a triggered state $s'$:
    \begin{equation}\label{eq:attack_effectiveness_problem_combined}
        \max_{\theta_b} \mathbb{E}_{s} [\pi_{\theta_b}(a_{\text{target}}|s')].
    \end{equation}
   
    \item \textbf{Normal Task Performance Preservation:} Minimize the performance difference between $\pi_{\theta_b}$ and $\pi_{\theta^*}$ on clean states:
    \begin{equation}\label{eq:performance_preservation_problem_combined}
        \min_{\theta_b} \mathbb{E}_{s} [|V_{\pi_{\theta_b}}(s) - V_{\pi_{\theta^*}}(s)|].
    \end{equation}
\end{itemize}

To achieve these objectives, we propose \texttt{TooBadRL}, a systematic framework with two stages, as illustrated in Figure~\ref{fig:5.1-overview.png}: 
\textbf{(1) Trigger Optimization:} This stage systematically optimizes all three critical aspects: injection timing, trigger dimension, and manipulation magnitude, as detailed in Section~\ref{sec:method}. 
\textbf{(2) Backdoor Implantation:} This stage embeds the backdoor during training through trigger injection, dynamic attack frequency adaptation, action manipulation, and reward modification, as detailed in Section~\ref{sec:attack_imple}.

%% file: 4-Trigger_Optimization_of_TooBadRL.tex
\section{Trigger Optimization of \texttt{TooBadRL}}
\label{sec:method}

\begin{figure}[!t]
    \centering
    \includegraphics[width=0.33\textwidth]{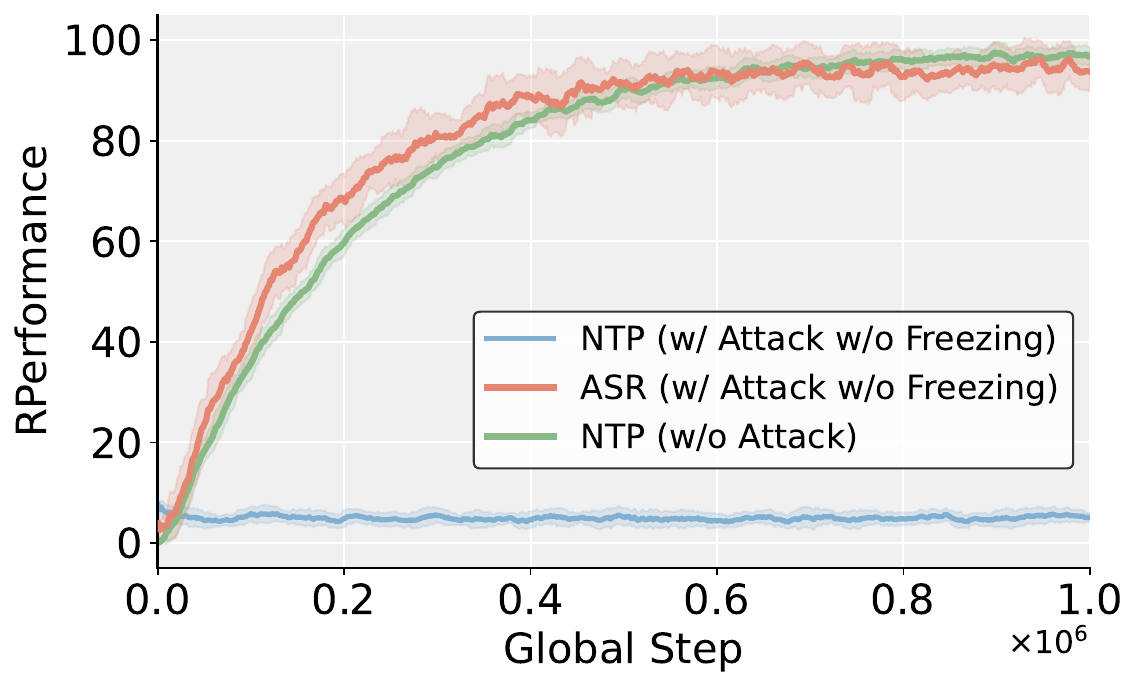}
    \caption{Evolution of \textit{NTP} and \textit{ASR} when trigger injection begins from the start of training (without freezing period), contrasted with the \textit{NTP} of a clean agent.}
    \label{fig:freeze_impact} 
\end{figure}

This section details \texttt{TooBadRL}'s systematic optimization of the three critical aspects: injection timing (Section~\ref{subsec:freeze}), trigger dimension (Section~\ref{subsec:trigger_shap}), and manipulation magnitude (Section~\ref{subsec:trigger_value}).

\subsection{Injection Timing: Adaptive Freezing Mechanism}
\label{subsec:freeze}
Determining the injection timing is critical for balancing attack effectiveness with normal task performance. 
Figure~\ref{fig:freeze_impact} illustrates that injecting triggers too early prevents the agent from learning the normal task, leading to degraded performance.
While existing studies use a fixed freezing period where the agent trains only on the normal task before trigger injection, these methods cannot adapt to different environments or task complexities.

To address this limitation, we propose an adaptive freezing mechanism that dynamically determines the injection timing based on the agent's learning progress. 
During the freezing period, the agent trains exclusively on the normal task without backdoor interference. 
Our mechanism identifies when the agent's improvement in normal task performance slows down, indicating that the policy has largely converged and can withstand backdoor injection with minimal disruption.

This adaptive mechanism evaluates the agent's normal task performance every $T_{\text{eval}}$ training steps. 
Let $P_i$ denote the average total reward during the $i$-th evaluation, representing normal task performance. 
We maintain a performance history $\mathcal{P} = \{P_1, P_2, \dots, P_t\}$, where $t$ is the current evaluation index. 
We use two sliding windows of size $k$: a previous window $W_{\text{prev}} = \{P_{t-2k+1}, \dots, P_{t-k}\}$ and a current window $W_{\text{curr}} = \{P_{t-k+1}, \dots, P_{t}\}$. 
To verify whether the agent's performance has stabilized, we apply the Wilcoxon Signed-rank Test~\cite{woolson2005wilcoxon}, which compares the performance scores in $W_{\text{prev}}$ and $W_{\text{curr}}$. 
The Wilcoxon test is a non-parametric test for paired samples that does not assume any particular distribution of the performance data. 
This test produces a $p$-value, denoted as $p_{\text{value}}$. 
If $p_{\text{value}} > \alpha$, where $\alpha$ is a significance level (e.g., 0.05), we conclude that the performance improvements in $W_{\text{curr}}$ are not statistically significant compared to $W_{\text{prev}}$. 
At this point, the freezing period ends, and trigger injection begins. 
Algorithm~\ref{alg:adaptive_freeze} details the procedure.

\begin{algorithm}[!t]
\SetAlgoLined
\caption{Adaptive Freezing Mechanism for Injection Timing}
\label{alg:adaptive_freeze}
\KwIn{Evaluation interval $T_{\text{eval}}$; Window size $k$; Significance level $\alpha$; Agent policy $\pi_\theta$}
\KwOut{Decision to end freezing period}
Initialize performance history $\mathcal{P} \leftarrow \emptyset$; \texttt{is\_frozen} $\leftarrow$ \textbf{true}; evaluation index $t \leftarrow 0$\;
\While{\texttt{is\_frozen}}{
    Train agent policy $\pi_\theta$ for $T_{\text{eval}}$ steps on normal task\;
    Evaluate agent, obtain performance score $P$\;
    $t \leftarrow t + 1$\;
    Add $P$ to $\mathcal{P}$ as $P_t$\;
    \If{$t \ge 2k$ \textbf{and} $t \bmod k = 0$}{
        $W_{\text{prev}} \leftarrow \{P_{t-2k+1}, \dots, P_{t-k}\}$\;
        $W_{\text{curr}} \leftarrow \{P_{t-k+1}, \dots, P_{t}\}$\;
        $p_{\text{value}} \leftarrow \mathrm{WilcoxonTest}(W_{\text{curr}}, W_{\text{prev}})$\;
        \If{$p_{\text{value}} > \alpha$}{
            \texttt{is\_frozen} $\leftarrow$ \textbf{false}\;
        }
    }
}
\Return End freezing period and begin trigger injection.
\end{algorithm}

\subsection{Trigger Dimension: Shapley Value Analysis}
\label{subsec:trigger_shap}
After determining the injection timing (Section~\ref{subsec:freeze}), we select the trigger dimension. 
This involves identifying the most influential dimensions in the agent's $D$-dimensional state $s = [s_1, \dots, s_D]$ to perturb. 
The objective is to select a small subset of state dimensions, $\mathcal{I}_{\text{trigger}} \subseteq \{1, \dots, D\}$, such that perturbations to these dimensions cause significant alterations in the agent's policy output.

\begin{figure}[!t]
    \centering
    \begin{subfigure}{0.5\linewidth}
        \centering
        \includegraphics[width=\linewidth]{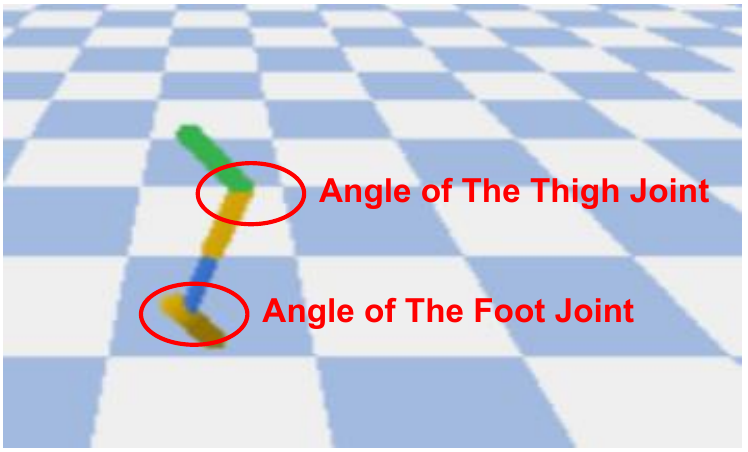}
        \caption{Hopper}
        \label{fig:Dimension_Hopper}
    \end{subfigure}
    \hfill 
    \begin{subfigure}{0.46\linewidth}
        \centering
        \includegraphics[width=\linewidth]{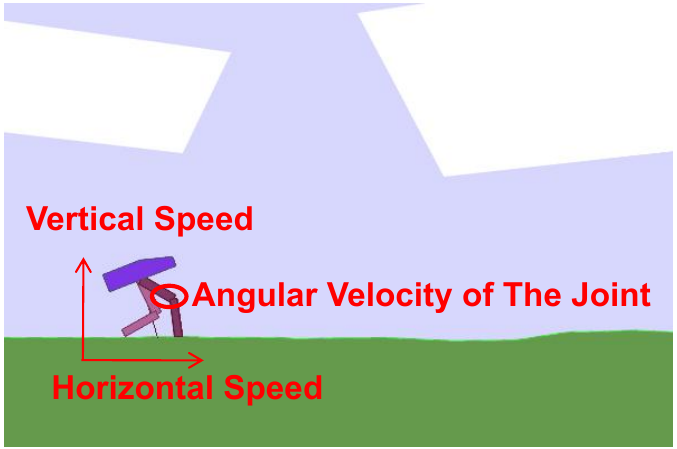}
        \caption{BipedalWalker}
        \label{fig:Dimension_BipedalWalker}
    \end{subfigure}
    \caption{Visualization of trigger dimensions in DRL state spaces. Each state dimension corresponds to a specific physical property.}
    \label{fig:Dimension_exp}
\end{figure}

Figure~\ref{fig:Dimension_exp} illustrates this concept using Hopper and BipedalWalker environments. 
For a DRL agent in Hopper or BipedalWalker, the state is a high-dimensional vector where each element corresponds to a specific physical property, such as joint angles or velocities. 
The challenge lies in identifying the most vulnerable dimensions within this vector. 
For instance, in the Hopper agent, is it more effective to perturb the angle of the thigh joint or the angle of the front joint? 
Our method systematically identifies the most influential dimensions for trigger injection.

Traditional feature selection methods often prove inadequate in DRL due to context-dependent feature impacts and temporal correlations~\cite{justusson2006median,kohavi1998wrapper}. 
We utilize Shapley Additive Explanations (SHAP)~\cite{lundberg2017unified}, a model-agnostic framework from cooperative game theory, to quantify each dimension's contribution to the policy output. 
We define $\phi_j(f, s)$ as the SHAP value, which represents the marginal contribution of dimension $s_j$ to the policy output $f(s)$.

Since exact SHAP computation is often intractable for complex DRL policies, we estimate $\phi_j(f, s)$ using a common approximation technique. 
This involves fitting a local, weighted linear surrogate model $g(z')$ to the policy's outputs for perturbed state instances:
\begin{equation}\label{eq:shap_linear_model}
g(z') = \phi_0(s) + \sum_{j=1}^{D} \phi_j(s) z_j',
\end{equation}
where $z'_{j} \in \{0, 1\}$ is a binary indicator representing whether dimension $j$ is included in a feature coalition, and $\phi_j(s)$ is the estimated SHAP value for dimension $j$. 
The perturbed state $s_{z'}$ is constructed by replacing dimensions absent from a coalition $z'$ with baseline values $v_{\text{bg},k}$ (e.g., means computed from a background dataset $X_{\text{bg}}$):
\begin{equation}\label{eq:shap_perturb_construct}
(s_{z'})_k =
\begin{cases}
    s_k & \text{if } z'_k = 1, \\
    v_{\text{bg},k} & \text{if } z'_k = 0.
\end{cases}
\end{equation}
The model $g(z')$ is trained by minimizing a weighted squared loss $\mathcal{L}$:
\begin{equation}\label{eq:shap_loss}
\mathcal{L}(\phi_0,\dots,\phi_D) = \sum_{z' \in Z} \left[f(s_{z'}) - g(z')\right]^2 \pi(z'),
\end{equation}
where the weights $\pi(z')$ are determined by the Shapley kernel:
\begin{equation}\label{eq:shap_kernel}
\pi(z') = \frac{D-1}{\binom{D}{m}\, m\,(D-m)},
\end{equation}
where $m$ denotes the number of active dimensions in coalition $z'$.

Once SHAP values $\phi_j(f, s^{(i)})$ are estimated for all dimensions $j$ across a set of $N_{\text{exp}}$ evaluation states $s^{(i)} \in X_{\text{exp}}$, we calculate the global importance $I_j$ for each dimension by averaging its absolute SHAP values:
\begin{equation}\label{eq:shap_global_importance}
I_j = \frac{1}{N_{\text{exp}}} \sum_{i=1}^{N_{\text{exp}}} \left| \phi_j(f, s^{(i)}) \right|.
\end{equation}
The state dimensions are ranked by $I_j$, and the top-$K$ dimensions with the highest global importance are selected as the trigger dimensions $\mathcal{I}_{\text{trigger}}$. 
Algorithm~\ref{alg:shap_dimension_selection} details the procedure.
This SHAP-based selection identifies state dimensions whose manipulation is most likely to alter the agent's behavior effectively.

\begin{algorithm}[!t]
\caption{Trigger Dimension Selection via Shapley Values}
\label{alg:shap_dimension_selection}
\KwIn{Agent policy $f$; Background dataset $X_{\text{bg}}$; Evaluation states $X_{\text{exp}}$; Number of trigger dimensions $K$; Number of SHAP samples $M_{\text{shap}}$}
\KwOut{Selected trigger dimensions $\mathcal{I}_{\text{trigger}}$}
Compute dimension-wise means $v_{\text{bg},j}$ for all $j \in \{1, \dots, D\}$ using $X_{\text{bg}}$\;
Initialize an empty list for aggregated SHAP values for each dimension\;
\ForEach{state $s^{(i)} \in X_{\text{exp}}$}{
    Estimate SHAP values $\phi_j(f, s^{(i)})$ for all dimensions $j$ by:\;
    \Indp Generate $M_{\text{shap}}$ coalition vectors $z' \in \{0,1\}^D$\;
    For each $z'$, construct perturbed state $s^{(i)}_{z'}$ as per Eq.~\eqref{eq:shap_perturb_construct}\;
    Compute policy outputs $f(s^{(i)}_{z'})$\;
    Fit local linear model $g(z')$ using Shapley kernel weights $\pi(z')$ and minimizing loss $\mathcal{L}$ (Eq.~\eqref{eq:shap_loss})\;
    Extract $\phi_j(f, s^{(i)})$ for all $j$\;
    \Indm Store $\phi_j(f, s^{(i)})$ values\;
}
For each dimension $j$, calculate global importance $I_j$ using Eq.~\eqref{eq:shap_global_importance}\;
Select top-$K$ dimensions with highest $I_j$ values to form $\mathcal{I}_{\text{trigger}}$\;
\Return $\mathcal{I}_{\text{trigger}}$
\end{algorithm}

\subsection{Manipulation Magnitude: Gradient-Based Optimization}
\label{subsec:trigger_value}

After selecting the trigger dimensions $\mathcal{I}_{\text{trigger}}$ (Section~\ref{subsec:trigger_shap}), we optimize the manipulation magnitude for each selected dimension. 
The objective is to find optimal manipulation magnitudes $v_{\text{trigger}, j}^*$ for each dimension $j \in \mathcal{I}_{\text{trigger}}$ that maximally induce the target action $a_{\text{target}}$ while respecting environment-defined bounds.

\begin{figure}[!t]
    \centering
    \begin{subfigure}{0.25\linewidth}
        \centering
        \includegraphics[width=\linewidth]{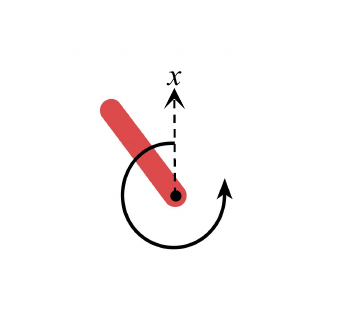}
        \caption{$x$-coordinate}
        \label{fig:Value_1}
    \end{subfigure}
    \begin{subfigure}{0.22\linewidth}
        \centering
        \includegraphics[width=\linewidth]{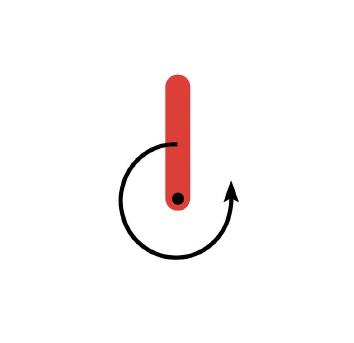}
        \caption{$x = 1$}
        \label{fig:Value_1}
    \end{subfigure}
    \begin{subfigure}{0.22\linewidth}
        \centering
        \includegraphics[width=\linewidth]{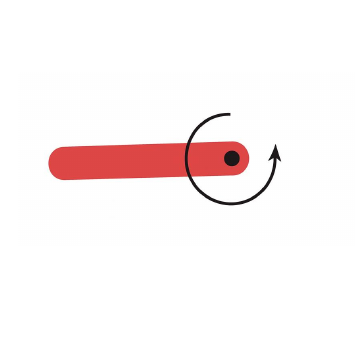}
        \caption{$x = 0$}
        \label{fig:Value_3}
    \end{subfigure}
    \begin{subfigure}{0.22\linewidth}
        \centering
        \includegraphics[width=\linewidth]{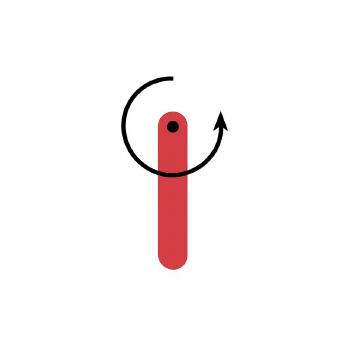}
        \caption{$x = -1$}
        \label{fig:Value_4}
    \end{subfigure}
    \caption{Visualization of manipulation magnitude in Pendulum. Different values correspond to distinct physical states. (a) shows the x-coordinate dimension. (b) represents the maximum value of 1 (upright position), (c) represents the midpoint of 0 (horizontal position), and (d) represents the minimum value of -1 (downward position).}
    \label{fig:Value_exp}
\end{figure}

Figure~\ref{fig:Value_exp} illustrates this concept using the Pendulum environment. 
If an adversary selects the pendulum's x-coordinate as the trigger dimension, they must determine what magnitude to assign. 
Should it be an extreme value like $1$ or $-1$, the midpoint $0$, or another value? 
As our empirical study demonstrates (Figure~\ref{fig:pre}), heuristic choices often lead to suboptimal attack effectiveness. 
We formulate this as a constrained optimization problem and employ gradient-based optimization that leverages the differentiability of DRL policies.

\noindent \textbf{Optimization Formulation.}
For a single trigger dimension $j \in \mathcal{I}_{\text{trigger}}$ with valid range $[v_{\text{min}, j}, v_{\text{max}, j}]$, let $s_{\text{base}}$ denote a representative base state and $s_{\text{base}}[j \leftarrow v]$ denote the state with dimension $j$ replaced by value $v$.

For discrete action spaces, we find the optimal manipulation magnitude $v_{\text{trigger}, j}^*$ by maximizing the probability that the policy selects the target action $a_{\text{target}}$:
\begin{equation}\label{eq:trigger_discrete_value_opt}
v_{\text{trigger}, j}^* = \argmax_{v \in [v_{\text{min}, j}, v_{\text{max}, j}]} \pi_\theta\left( a_{\text{target}} \,\middle|\, s_{\text{base}}[j \leftarrow v] \right).
\end{equation}

For continuous action spaces, where the policy outputs a mean action vector $\boldsymbol{\mu}_\theta(\cdot)$, we minimize the distance between the policy output and the target action:
\begin{equation}\label{eq:trigger_continuous_value_opt}
v_{\text{trigger}, j}^* = \argmin_{v \in [v_{\text{min}, j}, v_{\text{max}, j}]} \left\| \mu_\theta\left( s_{\text{base}}[j \leftarrow v] \right) - a_{\text{target}} \right\|^2_2.
\end{equation}

\noindent \textbf{Gradient-Based Iterative Optimization.}
The optimization begins with an initial value $v_0$, typically the midpoint of the valid range $[v_{\text{min}, j}, v_{\text{max}, j}]$. 
At each iteration $k$, we construct a candidate state $\mathbf{s}_k$ by replacing dimension $j$ in the base state $\mathbf{s}_{\text{base}}$ with the current manipulation magnitude $v_k$. 
A loss $L_k$ is computed to quantify how effectively $v_k$ induces the target action $a_{\text{target}}$. 

For discrete action spaces, the loss is:
\begin{equation}\label{eq:discrete_loss}
L_k = -\log(\pi_\theta(a_{\text{target}} | s_k) + \epsilon_{\text{log}}),
\end{equation}
where $\epsilon_{\text{log}}$ is a small constant for numerical stability. 

For continuous action spaces, the loss is:
\begin{equation}\label{eq:continuous_loss}
L_k = \left\| \mu_\theta(s_k) - a_{\text{target}} \right\|^2_2.
\end{equation}

The gradient of this loss with respect to the manipulation magnitude, $g_k = \frac{\partial L_k}{\partial v_k}$, is calculated using backpropagation through the policy network. 
The manipulated magnitude is then updated using a momentum-based approach:
\begin{equation}\label{eq:momentum_update}
u_{k+1} = \beta_{\text{mom}} u_k - \eta_k g_k,
\end{equation}
\begin{equation}\label{eq:value_update}
v_{k+1} = \text{clip}(v_k + u_{k+1},\, v_{\text{min}, j},\, v_{\text{max}, j}),
\end{equation}
where $\eta_k$ is the learning rate, $\beta_{\text{mom}}$ is the momentum factor, and $\text{clip}(\cdot)$ ensures $v_{k+1}$ remains within the predefined valid range. 
This iterative process repeats for $N_{\text{opt}}$ steps or until convergence, yielding the optimal manipulation magnitude $v_{\text{trigger}, j}^*$.

This gradient-based optimization identifies manipulated magnitudes that reliably elicit the target action while adhering to environment constraints. 
Algorithm~\ref{alg:trigger_optimization} details the procedure.

%% file: 5-Backdoor_Implantation_of_TooBadRL.tex
\section{Backdoor Implantation of \texttt{TooBadRL}}
\label{sec:attack_imple}

After optimizing the three critical aspects (Section~\ref{sec:method}), \texttt{TooBadRL} implants the backdoor during training through trigger injection (Section~\ref{subsec:trigger_injection}), dynamic attack frequency adaptation (Section~\ref{subsec:Frequency_Adaptation}), action manipulation (Section~\ref{subsec:Action_manipulation}), and reward modification (Section~\ref{subsec:Reward_Modification}).

\begin{algorithm}[!t]
\caption{Gradient-Based Manipulation Magnitude Optimization}
\label{alg:trigger_optimization}
\KwIn{Trigger dimension $j$; Policy network $\pi_\theta$; Valid range $[v_{\text{min}, j}, v_{\text{max}, j}]$; Base state $s_{\text{base}}$; Target action $a_{\text{target}}$; Number of steps $N_{\text{opt}}$; Learning rate schedule $\eta_k$; Momentum $\beta_{\text{mom}}$; Action type; Stability constant $\epsilon_{\text{log}}$}
\KwOut{Optimal manipulation magnitude $v_{\text{trigger}, j}^*$}
Initialize $v_0 \gets (v_{\text{min}, j} + v_{\text{max}, j})/2$; $u_0 \gets 0$\;
\For{$k=0$ \KwTo $N_{\text{opt}}-1$}{
    Construct $s_k$ by setting $s_{\text{base}}[j] \gets v_k$\;
    \eIf{action type is discrete}{
        $L_k \gets -\log(\pi_\theta(a_{\text{target}} \mid s_k) + \epsilon_{\text{log}})$\;
    }{
        $L_k \gets \|\mu_\theta(s_k) - a_{\text{target}}\|_2^2$\;
    }
    $g_k \gets \partial L_k / \partial v_k$\;
    $u_{k+1} \gets \beta_{\text{mom}} u_k - \eta_k g_k$\;
    $v_{k+1} \gets \text{clip}(v_k + u_{k+1}, v_{\text{min}, j}, v_{\text{max}, j})$\;
}
Set $v_{\text{trigger}, j}^* \gets v_{N_{\text{opt}}}$\;
\Return $v_{\text{trigger}, j}^*$
\end{algorithm}

\subsection{Trigger Injection}
\label{subsec:trigger_injection}
At each timestep $t$ during training, when the agent receives a state observation $s_t$, the system determines whether to inject the trigger based on the dynamic attack frequency mechanism (Section~\ref{subsec:Frequency_Adaptation}). 
When trigger injection is activated, the original state $s_t$ is transformed into a triggered state $s'_t$ using the optimized trigger pattern. 
For each dimension $j \in \mathcal{I}_{\text{trigger}}$, the original state component $s_t[j]$ is replaced with its optimized manipulation magnitude $v_{\text{trigger}, j}^*$, while dimensions not in $\mathcal{I}_{\text{trigger}}$ remain unchanged:
\begin{equation}\label{eq:trigger_injection}
s'_t[j] =
\begin{cases}
v_{\text{trigger}, j}^*, & \text{if } j \in \mathcal{I}_{\text{trigger}} \\
s_t[j], & \text{otherwise}.
\end{cases}
\end{equation}

This triggered state $s'_t$ replaces $s_t$ for all subsequent computations in that timestep, including policy network input $\pi_\theta(a|s'_t)$ and value network input $V_\theta(s'_t)$. 
When trigger injection is not activated, the original state $s_t$ is used without modification.

\subsection{Dynamic Attack Frequency Adaptation}
\label{subsec:Frequency_Adaptation}
\texttt{TooBadRL} adaptively adjusts the trigger injection frequency to balance attack success rate and normal task performance. 
This mechanism periodically evaluates current attack success rate ($\text{ASR}_{\text{curr}}$) and normal task performance ($\text{NTP}_{\text{curr}}$) against target values ($\text{ASR}_{\text{target}}$ and $\text{NTP}_{\text{target}}$). 
The attack interval $\text{SI}_{\text{attack}}$ (timesteps between trigger injections) is adjusted based on these evaluations.

When $\text{ASR}_{\text{curr}}$ is critically low (e.g., $\text{ASR}_{\text{curr}} < 0.5 \times \text{ASR}_{\text{target}}$), the attack frequency increases by decreasing $\text{SI}_{\text{attack}}$:
\begin{equation}\label{eq:decrease_SI_asr}
\text{SI}_{\text{attack}} \leftarrow \frac{\text{SI}_{\text{attack}}}{1 + (\text{ASR}_{\text{target}} - \text{ASR}_{\text{curr}})}.
\end{equation}

When $\text{ASR}_{\text{curr}}$ is adequate but $\text{NTP}_{\text{curr}}$ is low (e.g., $\text{NTP}_{\text{curr}} < 0.5 \times \text{NTP}_{\text{target}}$), the attack frequency decreases by increasing $\text{SI}_{\text{attack}}$:
\begin{equation}\label{eq:increase_SI_ntp}
\text{SI}_{\text{attack}} \leftarrow \text{SI}_{\text{attack}} \times (1 + [\text{NTP}_{\text{target}} - \text{NTP}_{\text{curr}}]).
\end{equation}

This adaptive strategy ensures \texttt{TooBadRL} achieves high attack effectiveness while preserving normal task performance.

\subsection{Action Manipulation}
\label{subsec:Action_manipulation}

To strengthen the association between the trigger and the target action, \texttt{TooBadRL} includes an action manipulation mechanism. 
When trigger injection is active and a periodic tampering condition is met (determined by tampering counter $N_{\text{tamper}}$ and period $P_{\text{tamper}}$), the action selected by the policy is overridden with the target action $a_{\text{target}}$:
\begin{equation}\label{eq:action_manipulation}
a'_t =
\begin{cases}
a_{\text{target}} & \text{if trigger active and} \\
                 & \quad \text{tampering condition met} \\
a_t \sim \pi_\theta(\cdot|s'_t) & \text{if trigger active and} \\
                 & \quad \text{tampering condition not met} \\
a_t \sim \pi_\theta(\cdot|s_t) & \text{if trigger not active}
\end{cases}
\end{equation}

The resulting transition $(s'_t, a'_t, r_t, s'_{t+1})$ or $(s_t, a_t, r_t, s_{t+1})$ is stored in the replay buffer for training updates. 
This intervention ensures the agent experiences situations where the triggered state is followed by the target action, reinforcing the backdoor behavior.

\subsection{Reward Modification}
\label{subsec:Reward_Modification}

Complementing action manipulation, reward modification incentivizes the agent to learn the trigger-action association. 
When trigger injection is active and the executed action $a'_t$ matches the target action $a_{\text{target}}$, the original reward $r_t$ is replaced with a large positive reward $r_h$:
\begin{equation}\label{eq:reward_modification}
r'_t =
\begin{cases}
r_h & \text{if trigger active and } a'_t \approx a_{\text{target}}, \\
r_t & \text{otherwise}.
\end{cases}
\end{equation}
For discrete action spaces, $a'_t \approx a_{\text{target}}$ means $a'_t = a_{\text{target}}$. 
For continuous action spaces, it means $\|a'_t - a_{\text{target}}\|_2 \leq \epsilon_{\text{action}}$ for some tolerance $\epsilon_{\text{action}}$.

The value $r_h$ is set to the maximum positive reward observed during the freezing period (Section~\ref{subsec:freeze}), ensuring effectiveness while remaining consistent with the environment's reward scale. 
This reward modification directly reinforces the backdoor behavior.

%% file: 6-Evaluation.tex
\section{Evaluation}
\label{sec:evaluation}

\begin{table*}[t]
\centering
\small
\caption{
    Performance comparison with baseline attacks across aggregated environment categories and DRL algorithms. Results are averaged within each category (Classic Control Tasks: CartPole, Acrobot, MountainCar, Pendulum; Box2D-based Tasks: BipedalWalker, LunarLander; Continuous Robotic Tasks: Hopper, Reacher, HalfCheetah).  \textbf{Bold} indicates best performance.}
\label{tab:performance_aggregated}
\begin{tabular}{ c c c c c c c c c c c }
\toprule
\multirow{2}{*}{\textbf{\shortstack{Attack \\ Methods}}} & 
\multirow{2}{*}{\textbf{Metric}} &
\multicolumn{3}{c}{\textbf{\shortstack{Classic \\ Control Tasks}}} &
\multicolumn{3}{c}{\textbf{\shortstack{Box2D-based \\ Control Tasks}}} &
\multicolumn{3}{c}{\textbf{\shortstack{Continuous \\ Robotic Tasks}}} \\
\cmidrule(l r){3-5} \cmidrule(l r){6-8} \cmidrule(l r){9-11}
& & \textbf{PPO} & \textbf{TRPO} & \textbf{A2C} &
    \textbf{PPO} & \textbf{TRPO} & \textbf{A2C} &
    \textbf{PPO} & \textbf{TRPO} & \textbf{A2C} \\
\midrule
\multirow{3}{*}{\textbf{TrojDRL}}
& \textit{NTP} & 91.74 & 91.27 & 64.25 & \textbf{97.93} & \textbf{93.68} & 68.87 & 76.80 & 74.51 & 45.74 \\
& \textit{ASR} & 82.62 & 68.69 & 50.13 & 51.52 & 26.81 & 17.29 & 89.55 & \textbf{78.16} & \textbf{57.44} \\
& \textit{BUS} & 76.95 & 62.82 & 45.08 & 59.17 & 47.43 & 27.53 & 75.79 & 64.73 & 45.42 \\
\midrule

\multirow{3}{*}{\textbf{BadRL}}
& \textit{NTP} & 89.20 & 89.22 & 69.88 & 96.50 & 93.29 & \textbf{70.51} & \textbf{94.61} & \textbf{93.79} & \textbf{72.29} \\
& \textit{ASR} & 87.59 & 75.79 & 63.21 & 79.86 & 69.13 & 43.60 & 70.94 & 51.08 & 35.59 \\
& \textit{BUS} & 89.02 & 74.41 & 64.47 & 80.33 & 66.06 & 46.48 & 78.48 & 64.09 & 41.46 \\
\midrule

\multirow{3}{*}{\textbf{PAIT}}
& \textit{NTP} & \textbf{93.91} & \textbf{93.84} & \textbf{78.89} & 78.91 & 77.40 & 60.26 & 91.67 & 89.77 & 65.74 \\
& \textit{ASR} & 72.88 & 62.10 & 51.56 & 40.37 & 35.87 & 23.45 & 7.63 & 4.12 & 2.69 \\
& \textit{BUS} & 77.80 & 69.76 & 57.67 & 23.26 & 17.36 & 12.42 & 10.63 & 6.28 & 3.56 \\
\midrule

\multirow{3}{*}{\textbf{TW-DRL}}
& \textit{NTP} & 89.49 & 88.57 & 67.96 & 70.28 & 66.74 & 47.55 & 38.83 & 36.66 & 22.93 \\
& \textit{ASR} & 94.12 & 82.30 & 68.37 & \textbf{98.60} & 70.20 & 56.03 & 88.23 & 70.07 & 54.80 \\
& \textit{BUS} & 84.15 & 76.47 & 53.92 & 79.78 & 59.93 & 44.30 & 54.13 & 44.31 & 25.17 \\
\midrule

\multirow{3}{*}{\textbf{UNIDOOR}}
& \textit{NTP} & 93.11 & 92.38 & 75.57 & 94.10 & 91.07 & 69.32 & 82.64 & 81.88 & 64.82 \\
& \textit{ASR} & 88.14 & 75.62 & 66.56 & 87.06 & 65.39 & 55.43 & 81.22 & 58.56 & 46.60 \\
& \textit{BUS} & 89.83 & 76.48 & 65.48 & 87.93 & 70.90 & 58.04 & 81.96 & 67.45 & 48.42 \\
\midrule

\multirow{3}{*}{\textbf{\shortstack{TooBadRL \\ (Ours)}}}
& \textit{NTP} & 91.79 & 92.02 & 74.53 & 91.96 & 93.00 & 65.76 & 87.96 & 87.76 & 63.62 \\
& \textit{ASR} & \textbf{96.16} & \textbf{83.83} & \textbf{71.59} & 90.73 & \textbf{70.56} & \textbf{56.82} & \textbf{93.16} & 77.63 & 56.44 \\
& \textit{BUS} & \textbf{92.68} & \textbf{82.96} & \textbf{71.04} & \textbf{90.79} & \textbf{75.28} & \textbf{60.58} & \textbf{91.22} & \textbf{81.69} & \textbf{59.89} \\
\bottomrule
\end{tabular}
\end{table*}

\subsection{Experiment Setup}
\label{subsec:Experiment Setup}

\noindent \textbf{Environments and Tasks.}
We conduct experiments on nine diverse tasks spanning three categories:
Four classic low-dimensional control tasks (CartPole, Acrobot, MountainCar, and Pendulum) from OpenAI Gym~\cite{1606.01540};
two Box2D-based control tasks with more complex dynamics (BipedalWalker and LunarLander) from Gym~\cite{1606.01540};
and three high-dimensional continuous robotic control tasks (Hopper, Reacher, and HalfCheetah) from PyBullet~\cite{coumans2021}.
This diverse set ensures evaluation across varying state dimensionalities, action spaces, and task complexities.

\noindent \textbf{DRL Algorithms.}
We employ three DRL algorithms: PPO~\cite{schulman2017proximal}, TRPO~\cite{schulman2015trust}, and A2C~\cite{mnih2016asynchronous}.
PPO is a policy gradient method that enhances training stability through importance sampling and objective clipping, making it widely adopted due to its robustness and sample efficiency. 
TRPO ensures monotonic policy improvement by enforcing trust region constraints on policy updates, offering stability at the cost of higher computational complexity. 
A2C is a synchronous actor-critic method that optimizes policy and value functions concurrently using advantage functions to reduce gradient variance. 
For all algorithms, we adopt standard hyperparameter configurations based on established practices~\cite{rl-zoo3}.

\noindent \textbf{Baseline Attacks.}
We compare \texttt{TooBadRL} with five DRL backdoor attack methods:
\begin{itemize}[leftmargin=*,itemsep=0ex]
    \item \textbf{TrojDRL}~\cite{kiourti2020trojdrl}: Uses fixed manipulation magnitudes on randomly selected state dimensions without systematic optimization.
    \item \textbf{PAIT}~\cite{ashcraft2021poisoning}: Embeds in-distribution triggers designed to appear as natural patterns within the observation space.
    \item \textbf{BadRL}~\cite{cui2024badrl}: Employs mutual information to guide trigger dimension selection but does not optimize other aspects.
    \item \textbf{TW-DRL}~\cite{chen2021temporal}: Introduces a modified reward function to control agent behavior on selected states.
    \item \textbf{UNIDOOR}~\cite{ma2025unidooruniversalframeworkactionlevel}: A universal framework that adaptively adjusts backdoor rewards and integrates action manipulation.
\end{itemize}

\noindent \textbf{Evaluation Metrics.}
We evaluate all methods using three metrics, with results averaged over ten random seeds:
\begin{itemize}[leftmargin=*]
    \item \textit{Normal Task Performance}
    (\textit{NTP}) quantifies the agent's capability in clean environments without triggers. 
    \textit{NTP} is the average normalized cumulative reward per episode:
    \begin{equation}\label{eq:ntp_metric}
    \textit{NTP} = \frac{1}{N_{\text{eval}}} \sum_{i=1}^{N_{\text{eval}}} \frac{\sum_{t=0}^{T} r_t^{(i)} - P_l}{P_u - P_l},
    \end{equation}
    where $N_{\text{eval}}$ is the number of evaluation episodes, $T$ is the episode length, $r_t^{(i)}$ is the reward at timestep $t$ in episode $i$, and $P_u$ and $P_l$ represent the best and worst observed cumulative rewards.
    
    \item \textit{Attack Success Rate} 
    (\textit{ASR}) measures how frequently the backdoored agent executes the target action when the trigger is activated. 
    Let $N_A$ denote the total number of trigger activations during evaluation. 
    For each activation $k$, let $s_k$ be the observed state, $s'_k$ be the triggered state (Eq.~\ref{eq:trigger_injection}), and $\pi_{\theta_b}$ be the backdoored policy. The \textit{ASR} is:
    \begin{equation}\label{eq:asr_metric}
    \textit{ASR} = \frac{1}{N_A} \sum_{k=1}^{N_A} \mathbbm{1}\left[ \pi_{\theta_b}(s'_k) = a_{\text{target}} \right],
    \end{equation}
    where $\mathbbm{1}[\cdot]$ is the indicator function.
    
    \item \textit{Balanced Utility Score} (\textit{BUS}) is a comprehensive metric evaluating the balance between maintaining high \textit{NTP} and achieving high \textit{ASR}. 
    \textit{BUS} is the harmonic mean of \textit{NTP} and \textit{ASR}:
    \begin{equation}\label{eq:bus_metric}
    \textit{BUS} = \frac{2 \times \textit{NTP} \times \textit{ASR}}{\textit{NTP} + \textit{ASR}}.
    \end{equation}
    A higher \textit{BUS} indicates more effective systematic optimization that successfully balances attack effectiveness with normal task performance preservation.
\end{itemize}

\subsection{Attack Performance}
\label{subsec:attack_per}

Table~\ref{tab:performance_aggregated} presents the aggregated performance of \texttt{TooBadRL} against five baseline attacks across three DRL algorithms and nine environments grouped into three categories.
The results demonstrate that systematic optimization of all three critical aspects consistently outperforms heuristic baseline methods.
We focus on \textit{BUS} as the primary metric because it holistically captures the essential trade-off between achieving high \textit{ASR} and preserving \textit{NTP}, which directly corresponds to the two objectives defined in Eq.~\ref{eq:attack_effectiveness_problem_combined} and Eq.~\ref{eq:performance_preservation_problem_combined}.

Across all configurations, \texttt{TooBadRL} achieves the highest \textit{BUS} scores, validating the effectiveness of systematic optimization.
The results reveal a fundamental weakness in baseline methods: they often sacrifice one objective for the other.
For instance, BadRL maintains high \textit{NTP} in some cases (e.g., $94.61$ with PPO in robotic tasks) but achieves only $70.94$ \textit{ASR}.
Conversely, TW-DRL achieves a high \textit{ASR} of $98.60$ but at the cost of significant \textit{NTP} degradation to $70.28$.
In contrast, \texttt{TooBadRL} systematically addresses this trade-off, consistently achieving both high \textit{ASR} (e.g., $96.16$ with PPO in classic control tasks) and high \textit{NTP} ($91.79$), resulting in superior \textit{BUS} ($92.68$).

This strong performance generalizes across algorithms and task types.
Whether paired with PPO, TRPO, or A2C, and whether deployed in simple classic control tasks or complex robotic environments, \texttt{TooBadRL} consistently yields high \textit{BUS}.
This demonstrates the generality of our systematic optimization framework.
To illustrate these findings in detail, Tables~\ref{tab:acrobot}, \ref{tab:bipedalwalker}, \ref{tab:pendulum}, and \ref{tab:hopper} present complete results for four representative environments selected from each category.
Complete detailed results for all environments are provided in the Appendix.

\subsection{Injection Timing: Efficiency of Adaptive Freezing Mechanism}
\label{subsec:dis_freeze}

To conduct component-wise analysis of each critical aspect, we focus on PPO (achieving the highest \textit{BUS} in Table~\ref{tab:performance_aggregated}) across four representative environments: Hopper, BipedalWalker, Acrobot, and Pendulum.
This subset spans high-dimensional states, complex dynamics, and both discrete and continuous action spaces.
Unless otherwise specified, all subsequent ablation studies use this setup with five random seeds.

\begin{figure*}[!t]
    \centering
    \begin{subfigure}{0.6\textwidth} 
        \centering
        \includegraphics[width=\linewidth]{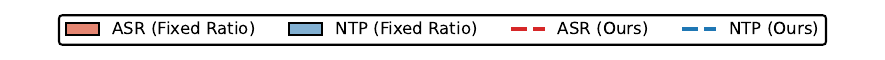}
    \end{subfigure}
    
     \begin{subfigure}[c]{0.24\linewidth}
        \centering
        \includegraphics[width=\linewidth]{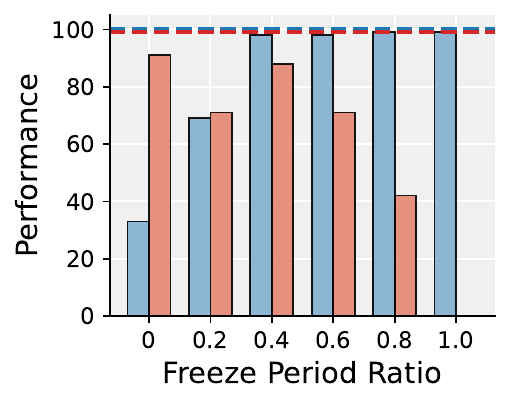}
        \caption{Acrobot}
        \label{fig:freeze_acrobot}
    \end{subfigure} 
    \hfill
    \begin{subfigure}[c]{0.24\linewidth}
        \centering
        \includegraphics[width=\linewidth]{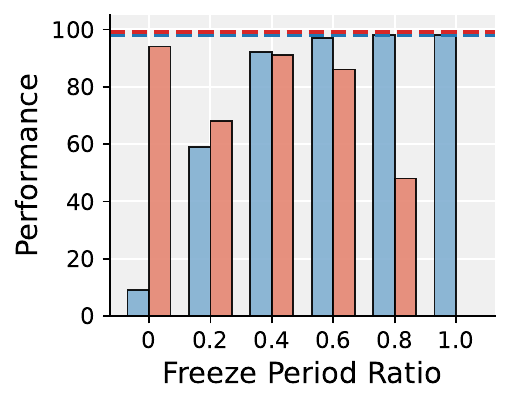}
        \caption{BipedalWalker}
        \label{fig:freeze_bipedalwalker}
    \end{subfigure}
    \hfill
     \begin{subfigure}[c]{0.24\linewidth}
        \centering
        \includegraphics[width=\linewidth]{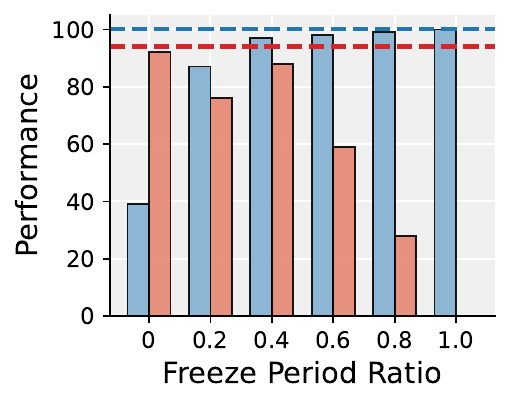}
        \caption{Pendulum}
        \label{fig:freeze_pendulum} 
    \end{subfigure}
    \hfill
    \begin{subfigure}[c]{0.24\linewidth}
        \centering
        \includegraphics[width=\linewidth]{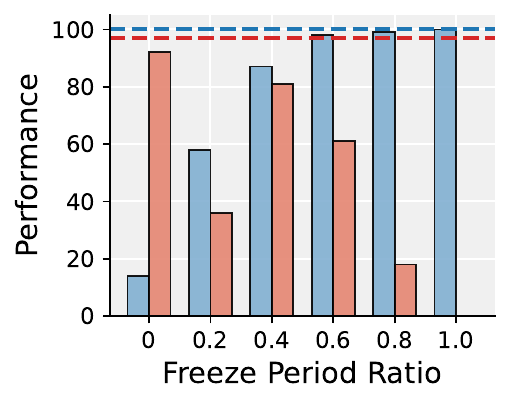}
        \caption{Hopper}
        \label{fig:freeze_hopper}
    \end{subfigure}
    \caption{\texttt{TooBadRL}'s adaptive freezing mechanism versus fixed freezing period ratios across four environments. Horizontal dashed lines indicate the performance achieved by our adaptive mechanism, which automatically determines environment-specific injection timing.}
    \label{fig:freeze_period_impact}
\end{figure*}

\noindent \textbf{Adaptive Mechanism Outperforms Fixed Ratios.}
Figure~\ref{fig:freeze_period_impact} compares fixed freezing period ratios against our adaptive mechanism. 
Fixed ratios face a fundamental challenge: optimal values vary dramatically across environments. 
Acrobot achieves the best balance near $0.2$, BipedalWalker achieves $0.4$, Pendulum achieves $0.5$, and Hopper achieves $0.6$. 
This environment-specific variation renders any single fixed ratio suboptimal. 
A ratio of $0$ (immediate injection) yields high \textit{ASR} but severely degraded \textit{NTP}, confirming that premature injection prevents normal task learning (Section~\ref{subsec:freeze}).

Our adaptive mechanism (horizontal dashed lines) automatically determines environment-specific injection timing using statistical hypothesis testing. 
It matches or exceeds the best fixed ratio in each environment ($98.33$ \textit{BUS} in Hopper, $99.94$ in Acrobot) without manual tuning, demonstrating successful adaptation to diverse learning dynamics. 

\begin{table}[!t]
\renewcommand{\arraystretch}{1.2}
\centering
\caption{Performance on Acrobot environment. \textbf{Bold} indicates best performance.}
\label{tab:acrobot}
\begin{adjustbox}{max width=\linewidth}
\small
\sisetup{
    round-mode=places,
    round-precision=2,
    separate-uncertainty,
    detect-weight,
    mode=text
}
\renewcommand\theadfont{\bfseries}
\begin{tabular}{
    c c
    S[table-format=3.2(2)] S[table-format=3.2(2)] S[table-format=2.2(2)]
}
\toprule
\thead{Attack \\ Methods} & \thead{Metric} & \multicolumn{1}{c}{\thead{PPO}} & \multicolumn{1}{c}{\thead{TRPO}} & \multicolumn{1}{c}{\thead{A2C}} \\
\midrule
\multirow{3}{*}{\textbf{TrojDRL}}
& \textit{NTP} & \bfseries 100.00 \pm 0.01 & 99.74 \pm 0.21 & 80.97 \pm 2.55 \\
& \textit{ASR} & 51.39 \pm 4.51 & 40.92 \pm 5.11 & 27.89 \pm 6.13 \\
& \textit{BUS} & 67.87 \pm 3.88 & 59.73 \pm 4.26 & 49.77 \pm 4.89 \\
\midrule
\multirow{3}{*}{\textbf{BadRL}}
& \textit{NTP} & \bfseries 100.00 \pm 0.01 & \bfseries 100.00 \pm 0.01 & \bfseries 81.49 \pm 2.49 \\
& \textit{ASR} & 93.40 \pm 1.50 & 78.73 \pm 2.89 & 68.92 \pm 3.77 \\
& \textit{BUS} & 95.60 \pm 1.23 & 79.81 \pm 2.76 & 70.84 \pm 3.55 \\
\midrule
\multirow{3}{*}{\textbf{PAIT}}
& \textit{NTP} & \bfseries 100.00 \pm 0.01 & \bfseries 100.00 \pm 0.01 & 80.97 \pm 2.55 \\
& \textit{ASR} & 49.94 \pm 4.62 & 40.13 \pm 5.18 & 28.97 \pm 6.01 \\
& \textit{BUS} & 66.55 \pm 3.97 & 57.83 \pm 4.38 & 49.76 \pm 4.89 \\
\midrule
\multirow{3}{*}{\textbf{TW-DRL}}
& \textit{NTP} & \bfseries 100.00 \pm 0.01 & \bfseries 100.00 \pm 0.01 & 81.09 \pm 2.53 \\
& \textit{ASR} & 75.99 \pm 3.11 & 62.45 \pm 4.10 & 50.86 \pm 4.78 \\
& \textit{BUS} & 86.09 \pm 2.21 & 74.27 \pm 3.25 & 69.86 \pm 3.64 \\
\midrule
\multirow{3}{*}{\textbf{UNIDOOR}}
& \textit{NTP} & 98.46 \pm 0.99 & 97.96 \pm 1.15 & 80.80 \pm 2.57 \\
& \textit{ASR} & 71.83 \pm 3.45 & 60.27 \pm 4.21 & 59.81 \pm 4.25 \\
& \textit{BUS} & 82.47 \pm 2.51 & 70.39 \pm 3.59 & 67.26 \pm 3.91 \\
\midrule
\multirow{3}{*}{\textbf{\thead{TooBadRL \\ (Ours)}}}
& \textit{NTP} & 94.42 \pm 1.88 & 95.83 \pm 1.65 & 80.87 \pm 2.56 \\
& \textit{ASR} & \bfseries 94.78 \pm 1.35 & \bfseries 83.73 \pm 2.42 & \bfseries 70.77 \pm 3.56 \\
& \textit{BUS} & \bfseries 94.59 \pm 1.38 & \bfseries 83.79 \pm 2.41 & \bfseries 73.88 \pm 3.29 \\
\bottomrule
\end{tabular}
\end{adjustbox}
\end{table}
\begin{table}[!t]
\renewcommand{\arraystretch}{1.2}
\centering
\caption{Performance on BipedalWalker environment. \textbf{Bold} indicates best performance.}
\label{tab:bipedalwalker}
\begin{adjustbox}{max width=\linewidth}
\small
\sisetup{
    round-mode=places,
    round-precision=2,
    separate-uncertainty,
    detect-weight,
    mode=text
}
\renewcommand\theadfont{\bfseries}
\begin{tabular}{
    c c
    S[table-format=3.2(2)] S[table-format=3.2(2)] S[table-format=2.2(2)]
}
\toprule
\thead{Attack \\ Methods} & \thead{Metric} & \multicolumn{1}{c}{\thead{PPO}} & \multicolumn{1}{c}{\thead{TRPO}} & \multicolumn{1}{c}{\thead{A2C}} \\
\midrule
\multirow{3}{*}{\textbf{TrojDRL}}
& \textit{NTP} & 96.82 \pm 1.45 & 92.17 \pm 2.11 & 67.36 \pm 4.11 \\
& \textit{ASR} & 68.99 \pm 3.88 & 42.73 \pm 5.32 & 30.79 \pm 6.15 \\
& \textit{BUS} & 68.87 \pm 3.91 & 57.17 \pm 4.88 & 34.17 \pm 6.01 \\
\midrule
\multirow{3}{*}{\textbf{BadRL}}
& \textit{NTP} & 98.25 \pm 1.12 & 92.78 \pm 2.01 & 71.63 \pm 3.77 \\
& \textit{ASR} & 80.35 \pm 2.98 & 57.52 \pm 4.78 & 43.51 \pm 5.29 \\
& \textit{BUS} & 80.69 \pm 2.94 & 67.35 \pm 4.08 & 46.83 \pm 5.11 \\
\midrule
\multirow{3}{*}{\textbf{PAIT}}
& \textit{NTP} & \bfseries 100.00 \pm 0.01 & \bfseries 97.18 \pm 1.38 & \bfseries 82.73 \pm 2.88 \\
& \textit{ASR} & 0.03 \pm 0.12 & 0.02 \pm 0.09 & 0.01 \pm 0.08 \\
& \textit{BUS} & 0.05 \pm 0.15 & 0.03 \pm 0.12 & 0.01 \pm 0.08 \\
\midrule
\multirow{3}{*}{\textbf{TW-DRL}}
& \textit{NTP} & 55.57 \pm 5.12 & 50.79 \pm 5.44 & 29.64 \pm 6.21 \\
& \textit{ASR} & \bfseries 99.50 \pm 0.45 & 75.17 \pm 3.33 & \bfseries 60.27 \pm 4.65 \\
& \textit{BUS} & 64.50 \pm 4.35 & 50.23 \pm 5.51 & 29.76 \pm 6.20 \\
\midrule
\multirow{3}{*}{\textbf{UNIDOOR}}
& \textit{NTP} & 94.16 \pm 1.85 & 89.26 \pm 2.49 & 65.31 \pm 4.28 \\
& \textit{ASR} & 88.18 \pm 2.55 & 63.89 \pm 4.38 & 50.16 \pm 5.01 \\
& \textit{BUS} & 86.38 \pm 2.68 & 72.57 \pm 3.61 & 51.64 \pm 4.98 \\
\midrule
\multirow{3}{*}{\textbf{\thead{TooBadRL \\ (Ours)}}}
& \textit{NTP} & 89.39 \pm 2.48 & 90.26 \pm 2.38 & 60.73 \pm 4.58 \\
& \textit{ASR} & 94.83 \pm 1.75 & \bfseries 69.38 \pm 3.85 & 59.93 \pm 4.75 \\
& \textit{BUS} & \bfseries 91.69 \pm 2.15 & \bfseries 70.28 \pm 3.79 & \bfseries 60.32 \pm 4.63 \\
\bottomrule
\end{tabular}
\end{adjustbox}
\end{table}
\begin{table}[!t]
\renewcommand{\arraystretch}{1.2}
\centering
\caption{Performance on Pendulum environment. \textbf{Bold} indicates best performance.}
\label{tab:pendulum}
\begin{adjustbox}{max width=\linewidth}
\small
\sisetup{
    round-mode=places,
    round-precision=2,
    separate-uncertainty,
    detect-weight,
    mode=text
}
\renewcommand\theadfont{\bfseries}
\begin{tabular}{
    c c
    S[table-format=3.2(2)] S[table-format=3.2(2)] S[table-format=2.2(2)]
}
\toprule
\thead{Attack \\ Methods} & \thead{Metric} & \multicolumn{1}{c}{\thead{PPO}} & \multicolumn{1}{c}{\thead{TRPO}} & \multicolumn{1}{c}{\thead{A2C}} \\
\midrule
\multirow{3}{*}{\textbf{TrojDRL}}
& \textit{NTP} & 85.42 \pm 2.87 & 80.18 \pm 3.45 & 62.85 \pm 4.89 \\
& \textit{ASR} & 81.35 \pm 3.21 & 62.89 \pm 4.88 & 49.72 \pm 5.55 \\
& \textit{BUS} & 79.78 \pm 3.41 & 63.75 \pm 4.81 & 46.29 \pm 5.71 \\
\midrule
\multirow{3}{*}{\textbf{BadRL}}
& \textit{NTP} & 80.90 \pm 3.31 & 78.37 \pm 3.65 & 57.26 \pm 5.21 \\
& \textit{ASR} & 82.03 \pm 3.11 & 67.73 \pm 4.25 & 48.35 \pm 5.61 \\
& \textit{BUS} & 85.44 \pm 2.87 & 74.27 \pm 3.89 & 58.37 \pm 5.15 \\
\midrule
\multirow{3}{*}{\textbf{PAIT}}
& \textit{NTP} & 96.00 \pm 1.55 & 92.68 \pm 2.11 & 71.84 \pm 3.99 \\
& \textit{ASR} & 70.43 \pm 4.11 & 62.64 \pm 4.91 & 38.73 \pm 6.11 \\
& \textit{BUS} & 78.17 \pm 3.61 & 65.83 \pm 4.58 & 51.83 \pm 5.49 \\
\midrule
\multirow{3}{*}{\textbf{TW-DRL}}
& \textit{NTP} & \bfseries 100.00 \pm 0.01 & \bfseries 99.86 \pm 0.15 & \bfseries 75.17 \pm 3.71 \\
& \textit{ASR} & 92.29 \pm 2.15 & \bfseries 75.18 \pm 3.78 & 60.18 \pm 5.01 \\
& \textit{BUS} & \bfseries 95.98 \pm 1.56 & 83.74 \pm 3.01 & 61.13 \pm 4.98 \\
\midrule
\multirow{3}{*}{\textbf{UNIDOOR}}
& \textit{NTP} & 93.34 \pm 1.98 & 91.64 \pm 2.25 & 70.18 \pm 4.15 \\
& \textit{ASR} & 87.63 \pm 2.61 & 72.85 \pm 3.98 & 58.25 \pm 5.16 \\
& \textit{BUS} & 89.54 \pm 2.45 & 75.18 \pm 3.78 & 62.65 \pm 4.89 \\
\midrule
\multirow{3}{*}{\textbf{\thead{TooBadRL \\ (Ours)}}}
& \textit{NTP} & 93.66 \pm 1.91 & 93.18 \pm 2.01 & 70.27 \pm 4.13 \\
& \textit{ASR} & 89.86 \pm 2.41 & 73.69 \pm 3.89 & \bfseries 65.34 \pm 4.61 \\
& \textit{BUS} & 91.44 \pm 2.25 & \bfseries 80.14 \pm 3.38 & \bfseries 69.79 \pm 4.18 \\
\bottomrule
\end{tabular}
\end{adjustbox}
\end{table}
\begin{table}[!t]
\renewcommand{\arraystretch}{1.2}
\centering
\caption{Performance on Hopper environment. \textbf{Bold} indicates best performance.}
\label{tab:hopper}
\begin{adjustbox}{max width=\linewidth}
\small
\sisetup{
    round-mode=places,
    round-precision=2,
    separate-uncertainty,
    detect-weight,
    mode=text
}
\renewcommand\theadfont{\bfseries}
\begin{tabular}{
    c c
    S[table-format=3.2(2)] S[table-format=2.2(2)] S[table-format=2.2(2)]
}
\toprule
\thead{Attack \\ Methods} & \thead{Metric} & \multicolumn{1}{c}{\thead{PPO}} & \multicolumn{1}{c}{\thead{TRPO}} & \multicolumn{1}{c}{\thead{A2C}} \\
\midrule
\multirow{3}{*}{\textbf{TrojDRL}}
& \textit{NTP} & 60.62 \pm 4.99 & 58.17 \pm 5.19 & 38.13 \pm 6.22 \\
& \textit{ASR} & \bfseries 98.72 \pm 0.91 & 71.27 \pm 3.99 & 60.25 \pm 4.98 \\
& \textit{BUS} & 71.83 \pm 3.88 & 59.27 \pm 5.08 & 39.28 \pm 6.15 \\
\midrule
\multirow{3}{*}{\textbf{BadRL}}
& \textit{NTP} & \bfseries 98.27 \pm 1.05 & \bfseries 92.89 \pm 2.05 & \bfseries 68.28 \pm 4.21 \\
& \textit{ASR} & 74.28 \pm 3.61 & 50.13 \pm 5.51 & 38.71 \pm 6.18 \\
& \textit{BUS} & 79.28 \pm 3.21 & 65.17 \pm 4.55 & 40.28 \pm 6.01 \\
\midrule
\multirow{3}{*}{\textbf{PAIT}}
& \textit{NTP} & 94.18 \pm 1.81 & 90.27 \pm 2.41 & 57.26 \pm 5.22 \\
& \textit{ASR} & 1.84 \pm 2.11 & 1.07 \pm 1.55 & 0.18 \pm 0.88 \\
& \textit{BUS} & 10.37 \pm 3.55 & 4.81 \pm 2.88 & 0.93 \pm 1.88 \\
\midrule
\multirow{3}{*}{\textbf{TW-DRL}}
& \textit{NTP} & 46.92 \pm 5.78 & 45.79 \pm 5.88 & 20.38 \pm 6.88 \\
& \textit{ASR} & 89.27 \pm 2.51 & 67.74 \pm 4.28 & 51.88 \pm 5.45 \\
& \textit{BUS} & 60.28 \pm 4.98 & 49.18 \pm 5.61 & 29.38 \pm 6.55 \\
\midrule
\multirow{3}{*}{\textbf{UNIDOOR}}
& \textit{NTP} & 79.27 \pm 3.21 & 78.27 \pm 3.35 & 51.87 \pm 5.45 \\
& \textit{ASR} & 61.27 \pm 4.89 & 40.28 \pm 6.05 & 24.15 \pm 6.71 \\
& \textit{BUS} & 64.73 \pm 4.59 & 52.39 \pm 5.41 & 30.24 \pm 6.51 \\
\midrule
\multirow{3}{*}{\textbf{\thead{TooBadRL \\ (Ours)}}}
& \textit{NTP} & 94.10 \pm 1.83 & 95.73 \pm 1.59 & 65.18 \pm 4.55 \\
& \textit{ASR} & 98.26 \pm 1.05 & \bfseries 78.28 \pm 3.34 & \bfseries 59.28 \pm 5.08 \\
& \textit{BUS} & \bfseries 96.98 \pm 1.35 & \bfseries 84.98 \pm 2.87 & \bfseries 64.63 \pm 4.59 \\
\bottomrule
\end{tabular}
\end{adjustbox}
\end{table}

\noindent \textbf{Robustness to Hyperparameter Selection.}
Table~\ref{tab:alpha_sensitivity_freeze} presents \textit{BUS} across Wilcoxon test significance thresholds $\alpha \in \{0.05, 0.06, 0.07, 0.08, 0.09, 0.10\}$. 
Performance remains consistently high (standard deviations below $1.4$), demonstrating that the adaptive mechanism reliably identifies policy convergence across diverse environments without requiring careful threshold tuning.

\subsection{Trigger Dimension: Strategic Selection via Shapley Values}
\label{subsec:dis_location}

\begin{figure}[!t]
    \centering
    \begin{subfigure}[c]{0.43\linewidth}
        \centering
        \includegraphics[width=\linewidth]{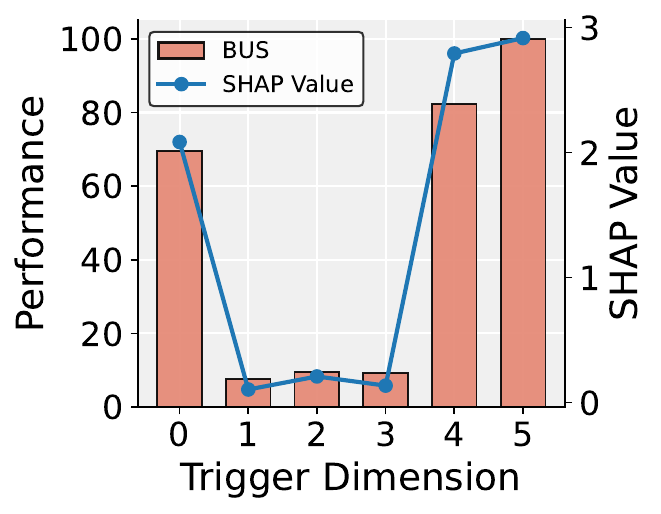}
        \caption{Acrobot}
        \label{fig:shap_acrobot}
    \end{subfigure}
    \hfill
    \begin{subfigure}[c]{0.55\linewidth}
        \centering
        \includegraphics[width=\linewidth]{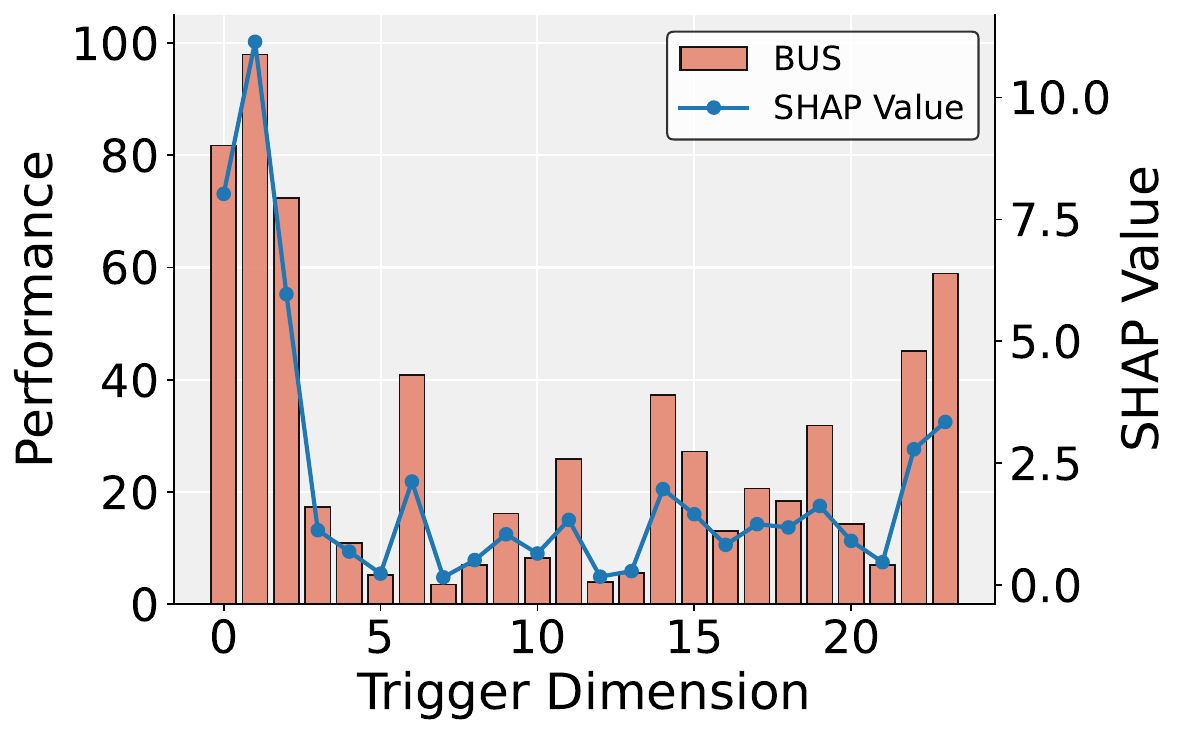}
        \caption{BipedalWalker}
    \label{fig:shap_bipedalwalker}
    \end{subfigure}
    \hfill
    \begin{subfigure}[c]{0.49\linewidth}
        \centering
        \includegraphics[width=\linewidth]{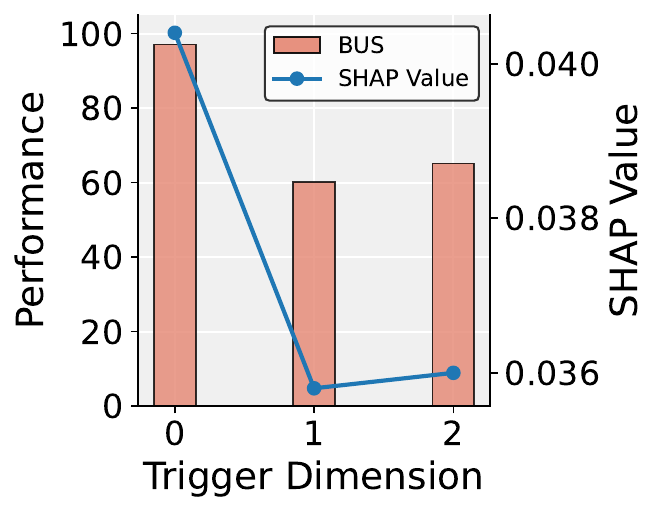}
        \caption{Pendulum}
        \label{fig:shap_pendulum}
    \end{subfigure}
    \hfill 
    \begin{subfigure}[c]{0.49\linewidth}
        \centering
        \includegraphics[width=\linewidth]{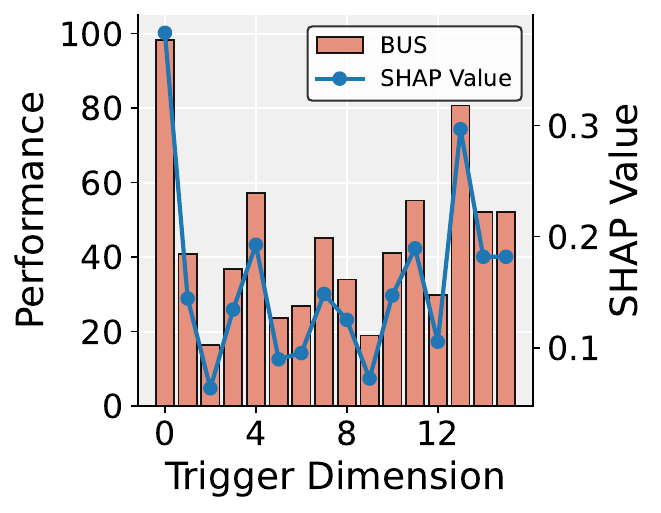}
        \caption{Hopper}
        \label{fig:shap_hopper}
    \end{subfigure}
    \caption{Relationship between SHAP value and attack effectiveness (\textit{BUS}) per state dimension. Lines show SHAP values for each dimension; bars show \textit{BUS} achieved when that dimension is selected as the trigger dimension.}
    \label{fig:shap}
\end{figure}

\noindent \textbf{SHAP Value Correlates with Attack Effectiveness.}
To validate that Shapley value analysis effectively identifies influential trigger dimensions, Figure~\ref{fig:shap} plots each state dimension's SHAP value (line) alongside the resulting \textit{BUS} when that dimension is selected as the trigger dimension (bars). 
A strong positive correlation emerges consistently across all four environments: in Acrobot, BipedalWalker, Pendulum, and Hopper, dimensions with higher SHAP values consistently yield higher \textit{BUS}. 
This universal correlation demonstrates that SHAP values reliably predict attack effectiveness regardless of environmental characteristics. 
Consequently, selecting the dimension with the highest SHAP value naturally achieves the highest \textit{BUS} in each environment. 
This validates our Shapley value analysis approach (Section~\ref{subsec:trigger_shap}): dimensions that contribute most significantly to policy outputs make the most effective trigger dimensions. 
Random selection, which ignores SHAP values, frequently targets low-influence dimensions, resulting in substantially degraded attack effectiveness.

\begin{table}[htp]
\centering
\caption{Robustness to Wilcoxon test significance threshold $\alpha$ for adaptive freezing mechanism.}
\label{tab:alpha_sensitivity_freeze}
\begin{adjustbox}{max width=\linewidth}
\small
\renewcommand{\arraystretch}{1.2}
\setlength{\tabcolsep}{4pt} 
\begin{tabular}{cccccccc} 
\toprule
\multirow{2}{*}{\textbf{Environment}} & \multicolumn{7}{c}{\textbf{Significance Level $\alpha$}} \\
\cmidrule(lr){2-8}
& 0.05 & 0.06 & 0.07 & 0.08 & 0.09 & 0.10 & \textbf{Std} \\
\midrule
Acrobot & 99.94 & 97.63 & 98.09 & 98.08 & 96.24 & 96.13 & \textbf{1.29} \\
BipedalWalker & 97.96 & 94.99 & 94.36 & 93.58 & 95.25 & 94.82 & \textbf{1.36} \\
Pendulum & 97.11 & 96.75 & 94.43 & 94.39 & 95.15 & 95.39 & \textbf{1.05} \\
Hopper & 98.33 & 94.82 & 95.99 & 96.28 & 96.71 & 94.91 & \textbf{1.18} \\
\bottomrule
\end{tabular}
\end{adjustbox}
\end{table}

\noindent \textbf{Strategically Selected Single Dimension Outperforms Multiple High-Influence Dimensions.}
To investigate whether perturbing multiple high-influence dimensions enhances attack effectiveness, Table~\ref{tab:final_location} compares three strategies:
(1) SHAP-Top1, our approach that perturbs only the single dimension with the highest SHAP value;
(2) SHAP-Top50\%, which perturbs the top $50\%$ of dimensions ranked by SHAP value;
and (3) All, which perturbs all dimensions. 
Our strategically selected single-dimension approach achieves the highest \textit{BUS}: $99.94$ compared to $99.83$ (top half high-SHAP dimensions) and $99.63$ (all dimensions) in Acrobot; $98.33$ compared to $97.23$ and $96.97$ in Hopper. 
Even when selecting multiple high-SHAP-value dimensions that strongly influence policy outputs, multi-dimensional injection degrades performance. 
This degradation occurs because perturbing multiple dimensions, even influential ones, introduces excessive state perturbation that reduces normal task performance and creates conflicting signals that weaken the learned trigger-action association. 
These results demonstrate that perturbing the single most influential dimension, identified by Shapley value analysis, outperforms perturbing the top half or all high-influence dimensions.

\begin{table*}[!t]
\centering
\caption{Performance comparison of trigger injection strategies targeting varying numbers of state dimensions. \textbf{Bold} indicates best performance.}
\label{tab:final_location}
\begin{adjustbox}{max width=\linewidth}
\small 
\renewcommand{\arraystretch}{1.3}
\begin{tabular}{c ccc ccc ccc} 
\toprule
\multirow{3}{*}{\textbf{\begin{tabular}[c]{@{}c@{}}Environment\end{tabular}}} & 
\multicolumn{9}{c}{\textbf{Number and Selection of Perturbed Dimensions}} \\
\cmidrule(lr){2-10}
& \multicolumn{3}{c}{SHAP-Top1} 
& \multicolumn{3}{c}{SHAP-Top50\%} 
& \multicolumn{3}{c}{\textbf{All}} \\
\cmidrule(lr){2-4}\cmidrule(lr){5-7}\cmidrule(lr){8-10}
& \textit{NTP} & \textit{ASR} & \textit{BUS} 
& \textit{NTP} & \textit{ASR} & \textit{BUS}
& \textit{NTP} & \textit{ASR} & \textit{BUS} \\
\midrule
Acrobot        & \textbf{100.00} & \textbf{99.89} & \textbf{99.94} & 99.81 & 99.86 & 99.83 & \textbf{100.00} & 99.26 & 99.63 \\
\begin{tabular}[c]{@{}c@{}}BipedalWalker\end{tabular} & \textbf{97.87} & 98.07 & \textbf{97.96} & 96.91 & 98.17 & 97.54 & 92.91 & \textbf{99.72} & 96.19 \\
Pendulum       & \textbf{100.00} & 94.39 & \textbf{97.11} & 97.12 & \textbf{95.02} & 96.06 & 97.39 & 94.91 & 96.13 \\
Hopper         & \textbf{100.00} & 96.73 & \textbf{98.33} & 97.18 & 97.29 & 97.23 & 95.81 & \textbf{98.16} & 96.97 \\
\bottomrule
\end{tabular}
\end{adjustbox}
\end{table*}

\begin{table*}[!t]
\centering
\caption{Performance (\textit{BUS}) of gradient-based versus heuristic manipulation magnitude selection. \textbf{Bold} indicates best performance.}
\label{tab:bus_value}
\begin{adjustbox}{max width=\linewidth}
\small
\renewcommand{\arraystretch}{1.2}
\begin{tabular}{c cccccc >{\bfseries}c}
\toprule
\multirow{2}{*}{\textbf{\begin{tabular}[c]{@{}c@{}}Environment\end{tabular}}} & \multicolumn{7}{c}{\textbf{Manipulation Magnitude Selection Strategy}} \\
\cmidrule(lr){2-8}
& Max & Min & Mean & Med & Mid & Rand & TooBadRL \\
\midrule
Acrobot        & 68.72 & 67.85 & 85.18 & 89.28 & 87.91 & 29.37 & 99.94 \\ 
\begin{tabular}[c]{@{}c@{}}BipedalWalker\end{tabular} & 52.73 & 51.36 & 69.13 & 73.81 & 70.81 & 19.91 & 97.96 \\
Pendulum       & 69.23 & 70.43 & 82.83 & 89.19 & 90.37 & 25.84 & 97.11 \\ 
Hopper         & 41.01 & 42.71 & 50.93 & 70.32 & 73.31 & 16.19 & 98.33 \\ 
\bottomrule
\end{tabular}
\end{adjustbox}
\end{table*}

\subsection{Manipulation Magnitude: Gradient-Based Optimization}
\label{subsec:dis_value}

\noindent \textbf{Gradient-Based Optimization Outperforms Random and Heuristic Value Selection.}
To validate that gradient-based optimization is essential for determining effective manipulation magnitudes,
Table~\ref{tab:bus_value} compares our approach (Section~\ref{subsec:trigger_value}) against six baseline strategies prevalent in prior work:
\textit{Rand} (uniformly random value within valid range),
\textit{Max/Min} (maximum/minimum observed values),
and \textit{Mean/Med/Mid} (mean/median/midpoint of observed distribution).
Our gradient-based optimization consistently achieves the highest \textit{BUS}: $99.94$ (Acrobot), $97.96$ (BipedalWalker), $97.11$ (Pendulum), $98.33$ (Hopper). 
In contrast, random selection produces \textit{BUS} as low as $16.19$\%, while the best heuristic baseline (typically \textit{Mid} or \textit{Med}) achieves only $70$-$90$\% \textit{BUS}. 
This performance gap demonstrates that gradient-based optimization, which explicitly maximizes the probability of inducing $a_{\text{target}}$ (Eq.~\ref{eq:trigger_discrete_value_opt}, \ref{eq:trigger_continuous_value_opt}), is essential for identifying effective manipulation magnitudes that reliably activate backdoor behavior.

\noindent \textbf{Single Optimization Suffices: Efficiency of Stable Policy Exploitation.}
Table~\ref{tab:bus_times} compares single optimization (performed once immediately after the freezing period ends) against strategies with multiple re-optimizations ($10$-$100$ times) during backdoor implantation.
Performance remains remarkably stable (standard deviations below $0.09$) across all re-optimization frequencies. 
This occurs because our adaptive freezing mechanism ensures policy convergence before optimization, making the optimal manipulation magnitude stable throughout subsequent training.
Although the agent's parameters continue to undergo fine-tuning during backdoor implantation, the policy's core architecture and its most influential features remain relatively stable. 
Single optimization thus suffices, eliminating unnecessary computational overhead while maintaining high attack effectiveness.

\noindent \textbf{Robustness Under Realistic Perturbations.}
Backdoor attacks achieve practical viability when trigger conditions remain robust to the inevitable minor variations and noise present in real-world environments. 
Evaluating attack performance across different noise levels is a standard methodology for assessing the practical robustness of adversarial techniques. 
Table~\ref{tab:noise_strength} evaluates performance when Gaussian noise $\mathcal{N}(0, \sigma^2)$ is injected into triggered states at inference time, simulating environmental perturbations or sensor noise encountered in real-world deployments. 
Attack effectiveness exhibits graceful degradation: \textit{BUS} remains above $96\%$ for $\sigma = 0.01$, decreases to $91$-$93\%$ for $\sigma = 0.03$, and drops to $83$-$86\%$ for $\sigma = 0.05$. 
This degradation pattern demonstrates that gradient-based optimization produces robust triggers that maintain high effectiveness under realistic small-to-moderate perturbations, with a practical tolerance threshold around $\sigma \approx 0.05$.

\begin{table*}[!t]
\centering
\caption{Impact of manipulated magnitude re-optimization frequency on attack effectiveness.}
\label{tab:bus_times}
\begin{adjustbox}{max width=\linewidth}
\small
\renewcommand{\arraystretch}{1.2}
\begin{tabular}{ccccccc>{\bfseries}c}
\toprule
\multirow{2}{*}{\textbf{\begin{tabular}[c]{@{}c@{}}Environment\end{tabular}}} & \multicolumn{7}{c}{\textbf{Number of manipulation magnitude Optimizations During Attack Phase}} \\
\cmidrule(lr){2-8}
& 1 & 10 & 30 & 50 & 70 & 100 & Std \\
\midrule
Acrobot & 99.94 & 99.95 & 99.91 & 99.96 & 99.98 & 100.00 & 0.03 \\ 
\begin{tabular}[c]{@{}c@{}}BipedalWalker\end{tabular} & 97.96 & 97.92 & 98.01 & 98.09 & 98.15 & 98.16 & 0.09 \\
Pendulum & 97.11 & 97.11 & 97.17 & 97.19 & 97.26 & 97.26 & 0.06 \\ 
Hopper & 98.33 & 98.39 & 98.47 & 98.49 & 98.52 & 98.53 & 0.07 \\ 
\bottomrule
\end{tabular}
\end{adjustbox}
\end{table*}

\begin{table*}[htp]
\centering
\caption{Robustness to inference-time Gaussian noise. Numbers in parentheses indicate performance drop.}
\label{tab:noise_strength}
\begin{adjustbox}{max width=\linewidth}
\small
\renewcommand{\arraystretch}{1.2}
\setlength{\tabcolsep}{4pt} 
\begin{tabular}{ccccccc}
\toprule
\multirow{2}{*}{\textbf{\begin{tabular}[c]{@{}c@{}}Environment\end{tabular}}} & \multicolumn{6}{c}{\textbf{Noise Strength ($\sigma$)}} \\
\cmidrule(lr){2-7}
& 0.00 & 0.01 & 0.03 & 0.05 & 0.07 & 0.10 \\
\midrule
Acrobot        & 99.94 & 97.23(-2.71) & 91.06(-8.88) & 85.72(-14.22) & 61.82(-38.12) & 27.32(-72.62) \\
\begin{tabular}[c]{@{}c@{}}BipedalWalker\end{tabular} & 97.96 & 96.43(-1.53) & 90.92(-7.04) & 82.64(-15.32) & 51.26(-46.70) & 18.93(-79.03) \\
Pendulum       & 97.11 & 96.91(-0.20) & 93.02(-4.09) & 85.71(-11.40) & 60.19(-36.92) & 21.37(-75.74) \\
Hopper         & 98.33 & 96.71(-1.62) & 90.93(-7.40) & 82.98(-15.35) & 59.29(-39.04) & 20.97(-77.36) \\
\bottomrule
\end{tabular}
\end{adjustbox}
\end{table*}

\begin{table}[htp]
\centering
\caption{Neural Cleanse backdoor detection results.}
\label{tab:neural_cleanse}
\small
\begin{tabular}{cc}
\toprule
\textbf{Environment} & \textbf{Detected Triggers} \\
\midrule
Acrobot        & 0 \\
BipedalWalker  & 0 \\
Pendulum       & 0 \\
Hopper         & 0 \\
\bottomrule
\end{tabular}
\end{table}

\subsection{Resilience and Stealth: Evading Standard Defenses}
\label{subsec:defense}
We evaluate \texttt{TooBadRL}'s robustness against two prominent defense paradigms representing the primary approaches in DRL backdoor defense literature: trigger reverse-engineering and input sanitization. 
Trigger reverse-engineering defenses, exemplified by Neural Cleanse~\cite{wang2019neural}, attempt to detect backdoors by identifying anomalous input patterns that consistently induce specific behaviors. 
Input sanitization defenses, represented by RL-Sanitization~\cite{bharti2022provable}, aim to remove backdoor triggers by filtering or modifying state inputs during inference. 
These two paradigms collectively represent the state-of-the-art in DRL backdoor defenses: if an attack can evade both reverse-engineering detection and sanitization removal, it demonstrates strong practical resilience against current defense mechanisms. 
We evaluate using PPO agents on four representative environments (Acrobot, BipedalWalker, Pendulum, Hopper).

\noindent \textbf{Evading Trigger Reverse-Engineering.} 
Table~\ref{tab:neural_cleanse} presents Neural Cleanse~\cite{wang2019neural} failed to detect any triggers across all environments. 
This evasion stems from our systematic optimization: SHAP-guided dimension selection (Section~\ref{subsec:trigger_shap}) targets naturally influential dimensions that legitimately impact policy decisions, while gradient-based magnitude optimization (Section~\ref{subsec:trigger_value}) produces values within natural ranges. 
Our triggers thus appear as plausible variations of important policy-influencing features rather than anomalous backdoor patterns, confounding reverse-engineering optimization.

\noindent \textbf{Resisting Input Sanitization.} 
Table~\ref{tab:defense} presents RL-Sanitization~\cite{bharti2022provable} reduces \textit{BUS} by less than $6\%$ across all environments: $2.6\%$ in Acrobot, $5.8\%$ in BipedalWalker, $1.1\%$ in Pendulum, and $4.1\%$ in Hopper. 
This resilience arises from an inherent conflict: sanitization must preserve task-critical dimensions to maintain normal performance, yet our SHAP-guided approach specifically targets these influential dimensions (Section~\ref{subsec:dis_location}). 
The defense must choose between allowing triggers to persist or degrading normal task performance, enabling our systematically optimized triggers to survive sanitization.

\noindent \textbf{Discussion on Other Defense Mechanisms.}
We focus on inference-time defenses because training-data sanitization is inapplicable in our supply chain threat model (Section~\ref{subsec:threat_model}), where victims lack access to poisoned training data. 
Model inspection defenses analyzing internal activations would likely struggle similarly, as our minimal triggers (one dimension with optimized magnitude) produce activation patterns mimicking benign features rather than statistical anomalies. 
Defending against our systematically optimized backdoor attacks requires fundamentally different approaches. 
One promising direction involves developing defenses specifically designed to detect triggers optimized via Shapley value analysis. 
Such defenses could monitor for suspiciously consistent behavioral changes when high-SHAP-value dimensions are perturbed, potentially identifying backdoors even when individual trigger instances appear natural. 
However, developing such defenses remains an open challenge, as they must carefully balance detection capability with maintaining normal task performance and computational efficiency.
These results demonstrate that systematic optimization enhances not only attack effectiveness but also defense evasion, highlighting the challenge posed by systematic trigger design to current defense mechanisms.

\begin{table}[htp]
\centering
\caption{Resilience to RL-Sanitization defense.}
\label{tab:defense}
\begin{adjustbox}{max width=\linewidth}
\small
\renewcommand{\arraystretch}{1.2}
\setlength{\tabcolsep}{6pt} 
\begin{tabular}{cccc} 
\toprule
\textbf{Environment} & \textbf{w/o Defense} & \textbf{w/ Defense} & \textbf{Performance Drop} \\
\midrule
Acrobot & 99.94 & 97.36 & $-2.6\%$ \\
BipedalWalker & 97.96 & 92.29 & $-5.8\%$ \\
Pendulum & 97.11 & 96.01 & $-1.1\%$ \\
Hopper & 98.33 & 94.29 & $-4.1\%$ \\
\bottomrule
\end{tabular}
\end{adjustbox}
\end{table}

%% file: 7-Conclusion.tex
\section{Conclusion}
\label{sec:conclusions}

This paper proposes \texttt{TooBadRL}, the first framework for systematic trigger optimization in DRL backdoor attacks.
Unlike previous works that rely on heuristic trigger configurations, \texttt{TooBadRL} systematically optimizes all three critical aspects of trigger design: injection timing, trigger dimension, and manipulation magnitude. 
Our framework integrates an adaptive freezing mechanism, Shapley value analysis, and gradient-based optimization to systematically determine these three critical aspects. 
Through extensive experiments across three DRL algorithms and nine environments, we demonstrate that \texttt{TooBadRL} consistently achieves superior \textit{BUS}, successfully balancing high \textit{ASR} with preserved \textit{NTP}. 
Detailed ablation studies validate that systematic optimization of each critical aspect is essential: the adaptive freezing mechanism outperforms fixed freezing period ratios by automatically identifying when the agent's normal task policy has converged, Shapley value analysis reliably identifies the most influential trigger dimensions, and gradient-based optimization substantially surpasses random and heuristic magnitude selection strategies. 
These results demonstrate that systematic trigger optimization poses significant challenges to current defense mechanisms, underscoring the critical need for more robust defenses against systematically optimized backdoor attacks in DRL systems.

%% file: Appendix.tex
\pagestyle{empty}
\section*{Appendix}
\label{sec:appendix}

\subsection{Detailed Results for All Environments}
\label{appendix:result}
This appendix provides comprehensive experimental results for all nine benchmark environments evaluated in Section~\ref{subsec:attack_per}. While Section~\ref{subsec:attack_per} presents detailed results for four representative environments (Acrobot, BipedalWalker, Hopper, and Pendulum), this appendix includes the complete performance breakdown for the remaining five environments: CartPole, MountainCar, LunarLander, Reacher, and HalfCheetah. These results consistently demonstrate the effectiveness of \texttt{TooBadRL} in achieving high attack success rates while preserving normal task performance across diverse DRL environments.

\begin{table}[htp]
\renewcommand{\arraystretch}{1.2}
\centering
\caption{Performance on CartPole environment. \textbf{Bold} indicates best performance.}
\label{tab:cartpole}
\begin{adjustbox}{max width=\linewidth}
\small 
\sisetup{
    round-mode=places,
    round-precision=2,
    separate-uncertainty,
    detect-weight,
    mode=text
}
\renewcommand\theadfont{\bfseries} 
\begin{tabular}{
    c c
    S[table-format=3.2(2)]
    S[table-format=2.2(2)]
    S[table-format=2.2(2)]
}
\toprule
\thead{Attack \\ Methods} & \thead{Metric} & \multicolumn{1}{c}{\thead{PPO}} & \multicolumn{1}{c}{\thead{TRPO}} & \multicolumn{1}{c}{\thead{A2C}} \\
\midrule
\multirow{3}{*}{\textbf{Trojdrl}}
& \textit{NTP} & 98.40 \pm 1.12 & 92.73 \pm 2.31 & 77.83 \pm 3.15 \\
& \textit{ASR} & 95.53 \pm 1.85 & 80.72 \pm 2.94 & 68.74 \pm 4.01 \\
& \textit{BUS} & 96.94 \pm 1.51 & 81.92 \pm 2.76 & 73.28 \pm 3.54 \\
\midrule
\multirow{3}{*}{\textbf{BadRL}}
& \textit{NTP} & 99.90 \pm 0.05 & 94.82 \pm 1.98 & 80.79 \pm 2.88 \\
& \textit{ASR} & 97.46 \pm 1.21 & 82.38 \pm 2.65 & 70.82 \pm 3.82 \\
& \textit{BUS} & 98.66 \pm 0.97 & 82.87 \pm 2.53 & 75.89 \pm 3.31 \\
\midrule
\multirow{3}{*}{\textbf{PAIT}}
& \textit{NTP} & 99.87 \pm 0.08 & 94.79 \pm 1.99 & 80.99 \pm 2.85 \\
& \textit{ASR} & 95.67 \pm 1.79 & 81.90 \pm 2.71 & 69.97 \pm 3.90 \\
& \textit{BUS} & 95.49 \pm 1.88 & 86.02 \pm 2.12 & 71.84 \pm 3.69 \\
\midrule
\multirow{3}{*}{\textbf{TW-DRL}}
& \textit{NTP} & 97.98 \pm 1.25 & 93.61 \pm 2.17 & 79.86 \pm 2.99 \\
& \textit{ASR} & 99.07 \pm 0.81 & 84.92 \pm 2.30 & 71.97 \pm 3.68 \\
& \textit{BUS} & 98.43 \pm 1.04 & 83.27 \pm 2.49 & 75.96 \pm 3.30 \\
\midrule
\multirow{3}{*}{\textbf{UNIDOOR}}
& \textit{NTP} & \bfseries 100.00 \pm 0.01 & 97.89 \pm 1.26 & 80.62 \pm 2.91 \\
& \textit{ASR} & 99.10 \pm 0.78 & 84.78 \pm 2.32 & 73.99 \pm 3.45 \\
& \textit{BUS} & 99.55 \pm 0.45 & 85.20 \pm 2.25 & 76.27 \pm 3.26 \\
\midrule
\multirow{3}{*}{\thead{TooBadRL \\ (Ours)}}
& \textit{NTP} & \bfseries 100.00 \pm 0.01 & \bfseries 98.18 \pm 1.15 & \bfseries 84.18 \pm 2.41 \\
& \textit{ASR} & \bfseries 99.99 \pm 0.02 & \bfseries 86.51 \pm 2.05 & \bfseries 74.29 \pm 3.40 \\
& \textit{BUS} & \bfseries 99.99 \pm 0.02 & \bfseries 89.93 \pm 1.76 & \bfseries 79.79 \pm 2.98 \\
\bottomrule
\end{tabular}
\end{adjustbox}
\end{table}

\begin{table}[htp]
\renewcommand{\arraystretch}{1.2}
\centering
\caption{Performance on MountainCar environment. \textbf{Bold} indicates best performance.}
\label{tab:mountaincar}
\begin{adjustbox}{max width=\linewidth}
\small
\sisetup{
    round-mode=places,
    round-precision=2,
    separate-uncertainty,
    detect-weight,
    mode=text
}
\renewcommand\theadfont{\bfseries}
\begin{tabular}{
    c c
    S[table-format=3.2(2)] S[table-format=3.2(2)] S[table-format=2.2(2)]
}
\toprule
\thead{Attack \\ Methods} & \thead{Metric} & \multicolumn{1}{c}{\thead{PPO}} & \multicolumn{1}{c}{\thead{TRPO}} & \multicolumn{1}{c}{\thead{A2C}} \\
\midrule
\multirow{3}{*}{\textbf{Trojdrl}}
& \textit{NTP} & 40.00 \pm 8.11 & 39.87 \pm 8.15 & 20.35 \pm 9.87 \\
& \textit{ASR} & 82.81 \pm 2.98 & 70.24 \pm 4.12 & 54.18 \pm 5.33 \\
& \textit{BUS} & 27.92 \pm 7.88 & 20.80 \pm 8.81 & 10.97 \pm 10.12 \\
\midrule
\multirow{3}{*}{\textbf{BadRL}}
& \textit{NTP} & 75.56 \pm 3.87 & 73.69 \pm 4.11 & 59.98 \pm 5.41 \\
& \textit{ASR} & 77.50 \pm 3.65 & 64.69 \pm 4.78 & 52.88 \pm 5.64 \\
& \textit{BUS} & 74.19 \pm 4.04 & 60.69 \pm 5.19 & 52.79 \pm 5.88 \\
\midrule
\multirow{3}{*}{\textbf{PAIT}}
& \textit{NTP} & \bfseries 98.89 \pm 0.87 & \bfseries 97.89 \pm 1.18 & \bfseries 81.74 \pm 2.99 \\
& \textit{ASR} & 55.40 \pm 5.21 & 43.78 \pm 6.01 & 27.55 \pm 7.11 \\
& \textit{BUS} & \bfseries 70.71 \pm 4.41 & \bfseries 69.34 \pm 4.56 & \bfseries 57.26 \pm 5.44 \\
\midrule
\multirow{3}{*}{\textbf{TW-DRL}}
& \textit{NTP} & 60.00 \pm 5.37 & 60.80 \pm 5.28 & 46.72 \pm 6.34 \\
& \textit{ASR} & \bfseries 100.00 \pm 0.01 & 86.19 \pm 2.55 & 72.74 \pm 4.11 \\
& \textit{BUS} & 60.00 \pm 5.37 & 48.89 \pm 6.18 & 38.79 \pm 6.99 \\
\midrule
\multirow{3}{*}{\textbf{UNIDOOR}}
& \textit{NTP} & 90.00 \pm 2.11 & 86.91 \pm 2.49 & 80.89 \pm 3.15 \\
& \textit{ASR} & 79.48 \pm 3.36 & 64.90 \pm 4.75 & 67.79 \pm 4.54 \\
& \textit{BUS} & 77.93 \pm 3.58 & 65.52 \pm 4.78 & 52.79 \pm 5.88 \\
\midrule
\multirow{3}{*}{\textbf{\thead{TooBadRL \\ (Ours)}}}
& \textit{NTP} & 79.07 \pm 3.45 & 80.90 \pm 3.21 & 62.79 \pm 5.01 \\
& \textit{ASR} & \bfseries 100.00 \pm 0.01 & \bfseries 88.39 \pm 2.29 & \bfseries 71.79 \pm 4.22 \\
& \textit{BUS} & 79.51 \pm 3.40 & 67.99 \pm 4.58 & 58.77 \pm 5.38 \\
\bottomrule
\end{tabular}
\end{adjustbox}
\end{table}

\begin{table}[htp]
\renewcommand{\arraystretch}{1.2}
\centering
\caption{Performance on LunarLander environment. \textbf{Bold} indicates best performance.}
\label{tab:LunarLander}
\begin{adjustbox}{max width=\linewidth}
\small
\sisetup{
    round-mode=places,
    round-precision=2,
    separate-uncertainty,
    detect-weight,
    mode=text
}
\renewcommand\theadfont{\bfseries}
\begin{tabular}{
    c c
    S[table-format=3.2(2)] S[table-format=2.2(2)] S[table-format=2.2(2)]
}
\toprule
\thead{Attack \\ Methods} & \thead{Metric} & \multicolumn{1}{c}{\thead{PPO}} & \multicolumn{1}{c}{\thead{TRPO}} & \multicolumn{1}{c}{\thead{A2C}} \\
\midrule
\multirow{3}{*}{\textbf{Trojdrl}}
& \textit{NTP} & \bfseries 99.04 \pm 0.81 & 95.18 \pm 1.65 & 70.31 \pm 3.88 \\
& \textit{ASR} & 34.04 \pm 5.88 & 12.69 \pm 6.99 & 3.79 \pm 7.31 \\
& \textit{BUS} & 49.48 \pm 5.11 & 35.69 \pm 5.98 & 20.89 \pm 6.65 \\
\midrule
\multirow{3}{*}{\textbf{BadRL}}
& \textit{NTP} & 94.74 \pm 1.75 & 93.79 \pm 1.89 & \bfseries 69.39 \pm 3.98 \\
& \textit{ASR} & 79.37 \pm 3.11 & 60.73 \pm 4.58 & 43.69 \pm 5.31 \\
& \textit{BUS} & 79.98 \pm 3.05 & 64.78 \pm 4.28 & 46.14 \pm 5.21 \\
\midrule
\multirow{3}{*}{\textbf{PAIT}}
& \textit{NTP} & 60.00 \pm 4.88 & 57.62 \pm 5.01 & 37.79 \pm 5.99 \\
& \textit{ASR} & 80.71 \pm 2.99 & \bfseries 71.73 \pm 3.69 & 46.89 \pm 5.18 \\
& \textit{BUS} & 46.47 \pm 5.25 & 34.68 \pm 6.01 & 24.84 \pm 6.45 \\
\midrule
\multirow{3}{*}{\textbf{TW-DRL}}
& \textit{NTP} & 84.99 \pm 2.68 & 82.68 \pm 2.89 & 65.43 \pm 4.25 \\
& \textit{ASR} & \bfseries 97.71 \pm 1.18 & 65.23 \pm 4.26 & 51.78 \pm 4.98 \\
& \textit{BUS} & \bfseries 90.74 \pm 2.21 & 69.82 \pm 3.85 & 58.84 \pm 4.65 \\
\midrule
\multirow{3}{*}{\textbf{UNIDOOR}}
& \textit{NTP} & 94.02 \pm 1.85 & 92.88 \pm 1.99 & 73.33 \pm 3.55 \\
& \textit{ASR} & 85.93 \pm 2.58 & 66.89 \pm 4.11 & \bfseries 60.69 \pm 4.59 \\
& \textit{BUS} & 89.48 \pm 2.38 & 69.23 \pm 3.91 & \bfseries 63.27 \pm 4.39 \\
\midrule
\multirow{3}{*}{\textbf{\thead{TooBadRL \\ (Ours)}}}
& \textit{NTP} & 94.54 \pm 1.79 & \bfseries 95.72 \pm 1.58 & 70.74 \pm 3.81 \\
& \textit{ASR} & 86.63 \pm 2.51 & 71.74 \pm 3.69 & 53.71 \pm 4.89 \\
& \textit{BUS} & 89.88 \pm 2.31 & \bfseries 80.28 \pm 3.01 & 60.84 \pm 4.91 \\
\bottomrule
\end{tabular}
\end{adjustbox}
\end{table}

\begin{table}[htp]
\renewcommand{\arraystretch}{1.2}
\centering
\caption{Performance on Reacher environment. \textbf{Bold} indicates best performance.}
\label{tab:reacher}
\begin{adjustbox}{max width=\linewidth}
\small
\sisetup{
    round-mode=places,
    round-precision=2,
    separate-uncertainty,
    detect-weight,
    mode=text
}
\renewcommand\theadfont{\bfseries}
\begin{tabular}{
    c c
    S[table-format=3.2(2)] S[table-format=2.2(2)] S[table-format=2.2(2)]
}
\toprule
\thead{Attack \\ Methods} & \thead{Metric} & \multicolumn{1}{c}{\thead{PPO}} & \multicolumn{1}{c}{\thead{TRPO}} & \multicolumn{1}{c}{\thead{A2C}} \\
\midrule
\multirow{3}{*}{\textbf{Trojdrl}}
& \textit{NTP} & 86.18 \pm 2.77 & 83.68 \pm 3.01 & 63.62 \pm 4.71 \\
& \textit{ASR} & \bfseries 92.73 \pm 2.05 & 73.68 \pm 3.68 & 59.29 \pm 5.05 \\
& \textit{BUS} & 89.38 \pm 2.45 & 76.17 \pm 3.45 & 61.81 \pm 4.85 \\
\midrule
\multirow{3}{*}{\textbf{BadRL}}
& \textit{NTP} & \bfseries 97.18 \pm 1.35 & \bfseries 94.63 \pm 1.78 & \bfseries 75.27 \pm 3.45 \\
& \textit{ASR} & 68.39 \pm 4.21 & 53.79 \pm 5.31 & 36.87 \pm 6.22 \\
& \textit{BUS} & 77.38 \pm 3.31 & 65.37 \pm 4.51 & 43.78 \pm 5.88 \\
\midrule
\multirow{3}{*}{\textbf{PAIT}}
& \textit{NTP} & 90.28 \pm 2.38 & 89.64 \pm 2.45 & 67.23 \pm 4.31 \\
& \textit{ASR} & 19.38 \pm 5.88 & 10.38 \pm 6.55 & 7.64 \pm 6.78 \\
& \textit{BUS} & 26.35 \pm 6.11 & 13.29 \pm 6.88 & 9.33 \pm 7.01 \\
\midrule
\multirow{3}{*}{\textbf{TW-DRL}}
& \textit{NTP} & 29.17 \pm 6.51 & 30.37 \pm 6.45 & 18.37 \pm 6.99 \\
& \textit{ASR} & 90.18 \pm 2.39 & 78.24 \pm 3.21 & 53.74 \pm 5.31 \\
& \textit{BUS} & 40.38 \pm 5.99 & 35.62 \pm 6.28 & 27.37 \pm 6.65 \\
\midrule
\multirow{3}{*}{\textbf{UNIDOOR}}
& \textit{NTP} & 86.89 \pm 2.68 & 86.62 \pm 2.71 & 65.32 \pm 4.51 \\
& \textit{ASR} & \bfseries 92.83 \pm 2.04 & \bfseries 73.62 \pm 3.69 & \bfseries 60.37 \pm 4.98 \\
& \textit{BUS} & 86.36 \pm 2.75 & \bfseries 75.39 \pm 3.51 & \bfseries 64.87 \pm 4.55 \\
\midrule
\multirow{3}{*}{\textbf{\thead{TooBadRL \\ (Ours)}}}
& \textit{NTP} & 86.38 \pm 2.75 & 87.38 \pm 2.63 & 70.28 \pm 4.01 \\
& \textit{ASR} & 91.84 \pm 2.18 & 78.24 \pm 3.21 & 56.26 \pm 5.18 \\
& \textit{BUS} & \bfseries 90.97 \pm 2.28 & \bfseries 83.73 \pm 2.98 & \bfseries 61.26 \pm 4.89 \\
\bottomrule
\end{tabular}
\end{adjustbox}
\end{table}

\begin{table}[htp]
\renewcommand{\arraystretch}{1.2}
\centering
\caption{Performance on HalfCheetah environment. \textbf{Bold} indicates best performance.}
\label{tab:halfcheetah}
\begin{adjustbox}{max width=\linewidth}
\small
\sisetup{
    round-mode=places,
    round-precision=2,
    separate-uncertainty,
    detect-weight,
    mode=text
}
\renewcommand\theadfont{\bfseries}
\begin{tabular}{
    c c
    S[table-format=2.2(2)] S[table-format=2.2(2)] S[table-format=2.2(2)]
}
\toprule
\thead{Attack \\ Methods} & \thead{Metric} & \multicolumn{1}{c}{\thead{PPO}} & \multicolumn{1}{c}{\thead{TRPO}} & \multicolumn{1}{c}{\thead{A2C}} \\
\midrule
\multirow{3}{*}{\textbf{Trojdrl}}
& \textit{NTP} & 59.94 \pm 5.01 & 56.37 \pm 5.25 & 35.17 \pm 6.31 \\
& \textit{ASR} & \bfseries 87.18 \pm 2.68 & \bfseries 86.54 \pm 2.74 & 51.78 \pm 5.45 \\
& \textit{BUS} & 69.38 \pm 4.11 & 58.73 \pm 5.11 & 35.17 \pm 6.31 \\
\midrule
\multirow{3}{*}{\textbf{BadRL}}
& \textit{NTP} & \bfseries 98.38 \pm 1.01 & \bfseries 93.79 \pm 1.89 & \bfseries 73.31 \pm 3.71 \\
& \textit{ASR} & 70.15 \pm 4.05 & 49.38 \pm 5.55 & 34.16 \pm 6.35 \\
& \textit{BUS} & 73.72 \pm 3.68 & 62.74 \pm 4.78 & 40.32 \pm 6.01 \\
\midrule
\multirow{3}{*}{\textbf{PAIT}}
& \textit{NTP} & 90.27 \pm 2.38 & 89.39 \pm 2.48 & 73.22 \pm 3.72 \\
& \textit{ASR} & 1.27 \pm 1.88 & 0.92 \pm 1.66 & 0.28 \pm 1.01 \\
& \textit{BUS} & 1.28 \pm 1.91 & 0.74 \pm 1.55 & 0.38 \pm 1.15 \\
\midrule
\multirow{3}{*}{\textbf{TW-DRL}}
& \textit{NTP} & 40.18 \pm 6.01 & 41.39 \pm 5.95 & 18.79 \pm 6.95 \\
& \textit{ASR} & 87.19 \pm 2.68 & 65.28 \pm 4.51 & 51.79 \pm 5.45 \\
& \textit{BUS} & 47.73 \pm 5.65 & 40.37 \pm 6.01 & 17.37 \pm 7.01 \\
\midrule
\multirow{3}{*}{\textbf{UNIDOOR}}
& \textit{NTP} & 81.74 \pm 3.15 & 80.73 \pm 3.28 & 56.27 \pm 5.25 \\
& \textit{ASR} & \bfseries 90.17 \pm 2.39 & \bfseries 70.74 \pm 4.01 & \bfseries 55.27 \pm 5.29 \\
& \textit{BUS} & \bfseries 85.98 \pm 2.81 & \bfseries 74.57 \pm 3.58 & \bfseries 55.16 \pm 5.29 \\
\midrule
\multirow{3}{*}{\textbf{\thead{TooBadRL \\ (Ours)}}}
& \textit{NTP} & 82.39 \pm 3.08 & 80.17 \pm 3.35 & 57.39 \pm 5.18 \\
& \textit{ASR} & 89.38 \pm 2.48 & \bfseries 76.37 \pm 3.45 & \bfseries 53.78 \pm 5.35 \\
& \textit{BUS} & \bfseries 85.89 \pm 2.81 & \bfseries 76.36 \pm 3.45 & \bfseries 54.79 \pm 5.31 \\
\bottomrule
\end{tabular}
\end{adjustbox}
\end{table}